\documentclass[
nofootinbib,           
amsmath,amssymb,
aps,
prresearch,
twocolumn,
floatfix, 
superscriptaddress]{revtex4-2}
\usepackage{color}
\usepackage{minitoc}
\usepackage{times}
\usepackage{cases}
\usepackage{gensymb}
\usepackage{booktabs}
\usepackage{graphicx}
\usepackage{dcolumn}
\usepackage{bm}
\usepackage[T1]{fontenc}
\usepackage{bibunits}
\usepackage{multirow}
\usepackage{hyperref}
\usepackage{natbib}

\hypersetup{
    colorlinks=true,
    linkcolor=red,      
    citecolor=blue,      
    urlcolor=blue       
}

\hypersetup{linktocpage,colorlinks=true,citecolor=blue}

\makeatletter
\AtBeginDocument{\let\LS@rot\@undefined}
\makeatother

\def \be {\begin{equation}}
\def \ee {\end{equation}}
\def \bee{\begin{equation*}}
\def \eee{\end{equation*}}

\def \dd  {{\rm d}}

\def \bvec {\left( \begin{array}{c}}
	\def \evec {\end{array}\right)}

\def \l1{\overline{1}} 

\newcommand{\nn}{\nonumber \\}

\newcommand{\moire }{moiré~}

\newcommand{\xfilll}[2][1ex]{%
	\dimen0=#2\advance\dimen0 by #1%
	\leaders\hrule height \dimen0 depth -#1\hfill%
}

\definecolor{deeppink}{rgb}{1.0, 0.08, 0.58}
\definecolor{brilliantrose}{rgb}{1.0, 0.33, 0.64}
\definecolor{darkorange}{rgb}{1.0, 0.55, 0.0}
\definecolor{darkpastelpurple}{rgb}{0.59, 0.44, 0.84}
\definecolor{red}{rgb}{1, 0.0, 0.0}

\DeclareFontFamily{U}{mathb}{\hyphenchar\font45}
\DeclareFontShape{U}{mathb}{m}{n}{
      <5> <6> <7> <8> <9> <10> gen * mathb
      <10.95> mathb10 <12> <14.4> <17.28> <20.74> <24.88> mathb12
      }{}
\DeclareSymbolFont{mathb}{U}{mathb}{m}{n}
\DeclareFontSubstitution{U}{mathb}{m}{n}
\DeclareMathSymbol{\circlearrowleft}       {3}{mathb}{"F6}
\DeclareMathSymbol{\circlearrowright}      {3}{mathb}{"F7}
\DeclareMathSymbol{\curvearrowleftright}   {3}{mathb}{"F2}

\newcommand\inter{\curvearrowleftright}

\newcommand{\intra}{{\mathchoice
    {
        \hspace{-0.7ex}
        {\mathrel{{\ooalign{\hss\raisebox{1.2ex}{
            \rotatebox{180}{$\circlearrowright$} 
            }\hss\cr\raisebox{1.2ex}{
            \rotatebox{180}{$\circlearrowleft$} 
            }}}}
        \hspace{-0.7ex}
        }
    }
    {
        \hspace{-0.7ex}
        {\mathrel{{\ooalign{\hss\raisebox{1.2ex}{
            \rotatebox{180}{$\circlearrowright$} 
            }\hss\cr\raisebox{1.2ex}{
            \rotatebox{180}{$\circlearrowleft$} 
            }}}}
        \hspace{-0.7ex}
        }
    }
    {
        \hspace{-1.5ex}
        {\mathrel{{\ooalign{\hss\raisebox{0.9ex}{
            \rotatebox{180}{
                \scalebox{0.7}{
                    $\circlearrowright$
                }
            } 
            }\hss\cr\raisebox{0.9ex}{
            \rotatebox{180}{
                \scalebox{0.7}{
                    $\circlearrowleft$
                }
            } 
            }}}}
        \hspace{-1.5ex}
        }
    }
    {
        \hspace{-1.3ex}
        {\mathrel{{\ooalign{\hss\raisebox{0.7ex}{
            \rotatebox{180}{
                \scalebox{0.6}{
                    $\circlearrowright$
                }
            } 
            }\hss\cr\raisebox{0.7ex}{
            \rotatebox{180}{
                \scalebox{0.6}{
                    $\circlearrowleft$
                }
            } 
            }}}}
        \hspace{-1.3ex}
        }
    }
}
} 

\newcommand{\nocontentsline}[3]{}
\let\origcontentsline\addcontentsline
\newcommand\stoptoc{\let\addcontentsline\nocontentsline}
\newcommand\resumetoc{\let\addcontentsline\origcontentsline}

\begin{document}
\begin{bibunit}[apsrev4-2]

\title{Quasi-two-dimensional spin helix and magnon-induced singularity in twisted bilayer graphene}

\author{Yung-Yeh Chang}
\affiliation{Institute of Physics, Academia Sinica, Taipei 11529, Taiwan}

\author{Kazuma Saito} 
\affiliation{Institute of Physics, Academia Sinica, Taipei 11529, Taiwan}
\affiliation{Department of Applied Physics, Tokyo University of Science, Katsushika, Tokyo 125-8585, Japan}

\author{Chen-Hsuan Hsu}
\affiliation{Institute of Physics, Academia Sinica, Taipei 11529, Taiwan}

\begin{abstract} 
Twisted bilayer graphene exhibits prominent correlated phenomena in two distinct regimes: a Kondo lattice near the magic angle, resembling heavy fermion systems, and a triangular correlated domain wall network under interlayer bias, akin to sliding Luttinger liquids previously introduced for cuprates. Combining these characteristics, here we investigate a system where interacting electrons in the domain wall network couple to localized spins. Owing to inter-domain-wall correlations, a quasi-two-dimensional spin helix phase within the localized spins emerges as a result of spatial phase coherence across parallel domain walls. Within the spin helix phase, magnons can induce a singularity, reflected in the scaling exponents of various correlation functions, accessible through electrical means and by adjusting the twist angle. We predict observable features in magnetic resonance and anisotropic paramagnetic spin susceptibility for the spin helix and the magnon-induced singularity, serving as experimental indicators of the interplay between the Kondo lattice and sliding Luttinger liquids. Integrating critical aspects of Luttinger liquid physics, magnetism, and Kondo physics in twisted bilayer graphene, our findings offer insights into similar correlated phenomena across a broad range of twisted van der Waals structures.
 
\end{abstract}

\maketitle

\stoptoc

\textit{Introduction.}
Twisted bilayer graphene (TBG) has emerged as a promising platform for exploring correlated phenomena. When the angle between the two layers are close to the magic angle, the Fermi velocity is dramatically suppressed~\cite{Lopes-TBG-PRL-2007,Bistritzer-PNAS-2011,castro-book-TBG,Vishwanath-2019-PRL,MYChou-PRB-2024}, resulting in the formation of quasiflat energy bands and a significant enhancement in the density of states. Consequently, interactions between electrons become significant as compared to the bandwidth,  leading to various correlated quantum states~\cite{Andrei-NatMat-2020, Andrea-2020-NaPhys-moire, Yazdani-2024-moire-review, andrei-nat-rev-2021-moire}, such as superconductivity~\cite{Cao-2018-Nature-SC-TBG}, correlated insulating states~\cite{Cao-2016-PRL-correlInsulator-TBG,Cao-2018-Nature-correlaInsulator-TBG}, 
strange metals~\cite{Young-NatPhys-strangemetal-TBG,Cao-strangemetal-2020-PRL},  orbital ferromagnetism~\cite{orbital-ferro-2020-nat-Andrea,Goldhaber-Gordon-Science-orb-ferro-moire,Serlin:2020,Goldhaber-Gordon-orb-ferro-Nanolett}, and nematic order~\cite{Cao-Nematic-moire-science,moire-nematic-NatPhys-2022}. 
The many-body correlations are believed to originate from the strongly localized wavefunctions at AA-stacking regions~\cite{Koshino-PRX-2018,PHChun-PRX-2018,Yazdani-localAA-STM-TBG,Vishwanath-2019-PRL,Vafek-PRX-2018,Yazdani-cascade-2020,Saito-TBG-Pomeranchuk,Rozen-Nature-TBG-Pomeranchuk,EvaAndrei-TBG-NatComm-2021,XDai-PRB-2022}.
Furthermore, recent investigations revealed that magic angle TBG shows characteristics akin to those of heavy fermion compounds, where localized moments develop near AA-stacking regions and couple to conduction electrons through spin-exchange coupling~\cite{SB-TBG-Kondo-PRL,Chou-TBG-Kondo-2023,Haoyu-PRL-2023-Kondo-TBG,Haoyu-2023-PRL-Symm-Kondo-TBG,Haoyu-arxiv-HF-TBG,Calugaru-arxiv-Seeback-MATBG,XDai-PRB-2022,Liam-Coleman:2023}, thereby providing an additional perspective on Kondo-lattice systems.

Remarkably, correlated phenomena in TBG can be achieved without relying on specific twisted structures. In particular, upon applying perpendicular electric fields, domain walls separating the AB- and BA-stacking areas are known to host gapless modes~\cite{San-Jose-TBG-network,Alden2013,Ju2015,Yin2016,Efimkin-2018-TBG-network,Koshino-moireDW-2020,moire-DW-Ensslin,LeRoy-moire-DW-exp,PhilipKim-moire-DW,Xu-moireDW-GiantOscil,Carr-DW-hetero,Peeters-PRMater,Fleischmann-NanoLett-2020,Ren-PRB-Metallic-network,Walet-2020,Verbakel-PRB}, 
which form a two-dimensional (2D) triangular quantum network. This network exhibits correlated phenomena and high tunability~\cite{HCWang-TBG}, with electrically adjustable parameters such as Fermi velocity, bandwidth and interaction strengths of the domain wall modes, which can further control the instability of the network towards various orders. 
This distinct regime highlights that TBG can host correlated phenomena across a broad range of configurations, independent of precise stacking or magic-angle conditions.
   
\begin{figure*}[ht]
    \centering    
    \includegraphics[width=0.95\textwidth]{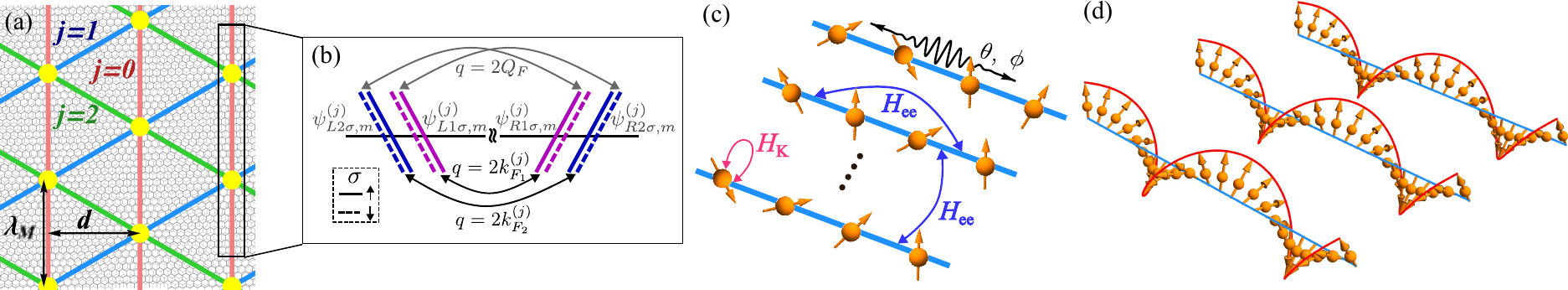}
    \caption{(a) TBG domain wall network formed by three arrays (labeled by $j = \lbrace 0,1,2 \rbrace$) of parallel domain walls (red, blue, and green lines), each rotated by $120^{\degree}$ relative to the others. The domain walls are indexed by $m \in [1,N_{\rm{dw}}]$ with the number $N_{\rm dw}$ of domain walls per array. 
    Here, $\lambda_M$ represents the \moire wavelength, and $d$ the distance between two adjacent domain walls. (b) Within a single domain wall, there are eight low-energy modes with Fermi wave vectors $k^{(j)}_{F_{\delta},m}$ and represented by fermion fields, $\psi^{(j)}_{\ell \delta \sigma,m}$ with the spin  $\sigma \in \left\lbrace\uparrow (\rm{solid}), \downarrow  (\rm{dashed}) \right\rbrace$, propagation directions $\ell\in \left\lbrace R , L \right\rbrace$, and  branches $\delta \in \lbrace 1  (\rm{blue}), 2 (\rm{purple}) \rbrace $.  
    The arrows indicate backscattering processes with  momentum transfers $q \in \left\lbrace  2k_{F_1,m}^{(j)}, \;  2k_{F_2,m}^{(j)} , \; 2Q_F \equiv k_{F_1,m}^{(j)} + k_{F_2,m}^{(j)}   \right\rbrace$.
   (c) Schematic of the $e$-$e$ interactions in parallel domain walls (blue lines) and spin-exchange interaction $H_{\rm K}$ with localized moments (represented as an orange arrow and sphere). The former includes contributions within and between parallel domain walls, leading to sliding Luttinger liquids characterized by bosonic fields (wavy curve) corresponding to the fermion fields in Panel~(b).  
   (d) Sketch of spatially phase-coherent spin helices on three adjacent domain walls.
}
    \label{fig:schematic}
\end{figure*}

Motivated by these discoveries, here we explore a system that combines two distinct characteristics in TBG, including correlated network and the coupling between itinerant carriers and localized magnetic moments [Figs.~\ref{fig:schematic}(a)-\ref{fig:schematic}(c)]. The coupling between the interacting electrons and localized spins in such quasi-2D correlated networks has remained insufficiently explored. Specifically, we examine a system comprising interacting electrons that traverse the domain walls, coupling to localized moments distributed on the graphene layers through Kondo-type interaction; see Figs.~\ref{fig:schematic}(a) and \ref{fig:schematic}(c). The first entity forms a triangular domain wall network~\cite{castro-TBG-network-PRB,Hsu:2023}, extending coupled wire models from earlier studies on cuprates~\cite{Emery-sLL-2000,Carpentier-sLL-2001,Mukhopadhyay-sLL-PRB-2001,Mukhopadhyay-csLL-PRB}. The localized moments could be introduced through magnetic adatoms~\cite{F-adatom-graphene,Vojta-RPP-graphene-Kondo} or nuclear spins via isotope engineering~\cite{Fischer2009,Churchill2009-naturephysics}. The electrons mediate a spatially oscillating indirect coupling between these moments within and across parallel domain walls, leading to helical magnetic ordering of the localized moments at sufficiently low temperatures. In contrast to isolated one-dimensional systems~\cite{Braunecker-2009,Braunecker-PRL-2009-C13,Meng-2013-PRB-2-subband,Jelena-2013-PRL-RKKY-TSC,Meng-strip-EuroPhys,CHH-CNT-helix,Hsu:2020}, the spatial phase coherence developed in parallel domain walls leads to the formation of a quasi-2D spin helix; see Fig.~\ref{fig:schematic}(d). Within the spin helix phase, we identify a magnon-induced singularity, reflected in the scaling exponents of various correlation functions and the carrier velocity. This singularity is accessible not only by electrical means but also by varying the twist angle. We further predict observable features of the spin helix and magnon-induced singularity in magnetic resonance and paramagnetic susceptibility, providing an electrically tunable platform for the interplay between correlated electrons and localized spins.

{\it Model.}
Our Hamiltonian $H =H_{{\rm ee}} + H_{{\rm K}}$ includes the electronic subsystem, $H_{{\rm ee}} $, and their coupling, $H_{{\rm K}}$, to localized moments in TBG with a small but finite twist angle. Incorporating an interlayer bias in the continuum model~\cite{Bistritzer-PNAS-2011,Efimkin-2018-TBG-network}, one can compute the density profile of the conduction electrons confined in the domain walls in order to construct the network model~\cite{HCWang-TBG}. 
Using the standard bosonization~\cite{Giamarchi2003}, the electron subsystem, including both the kinetic energy and electron-electron ($e$-$e$) interactions, is described by
\begin{align}
    H_{{\rm ee}}  = & \frac{1}{N_{\rm dw}} \sum_{j, q_{\perp}}\sum_{\nu,P}\int\frac{\hbar dr}{2\pi}\left[\frac{v_{\nu P}(q_{\perp})}{\tilde{K}_{\nu P}(q_{\perp})}\left|\partial_{r}\phi_{\nu P,q_{\perp}}^{j}(r)\right|^{2}\right. \nn
    & \qquad \left. +v_{\nu P}(q_{\perp})\tilde{K}_{\nu P}(q_{\perp})\left|\partial_{r}\theta_{\nu P,q_{\perp}}^{j}(r)\right|^{2}\right]\;,
    \label{eq:Hee}
\end{align}
with the momentum component $\hbar q_{\perp}$ perpendicular to the domain walls. 
The bosonic fields $\phi$ and $\theta$ in Eq.~(\ref{eq:Hee}) are labeled as the charge/spin  $\nu \in  \lbrace c, s\rbrace $ and symmetric/antisymmetric $P\in \lbrace S,A\rbrace$ sectors, corresponding to the linear combination of the eight gapless modes per domain wall in Fig.~\ref{fig:schematic}(b), and satisfy the commutation relation, 
\begin{align}
    \left[\phi_{\nu P,q_\perp}^{j}(r),\theta_{\nu^{\prime}P^{\prime},q_\perp^{\prime}}^{j^{\prime}}(r^{\prime})\right] =\frac{i\pi}{2}& \delta_{jj^{\prime}}\delta_{\nu\nu^{\prime}} \delta_{PP^{\prime}}\delta_{q_\perp, -q_\perp^{\prime}}\nn
    & \times \text{sgn}(r^{\prime}-r) \; . 
\end{align}
Eq.~\eqref{eq:Hee} is governed by $q_\perp$-dependent  functions, $\tilde{K}_{\nu P}$ and $v_{\nu P}$, serving as the effective interaction strength and carrier velocity generalized for the network. Since screened Coulomb interactions only enter the charge-symmetric sector ($\nu P = cS$)~\cite{HCWang-TBG}, for $\nu P \neq cS$ we take $\tilde{K}_{\nu P} =1$ with $v_{\nu P}$ given by the domain wall velocity $v_{\rm dw}$. We further assume periodic boundary condition perpendicular to the domain walls within each array~\cite{Emery-sLL-2000,Carpentier-sLL-2001,Mukhopadhyay-sLL-PRB-2001,Mukhopadhyay-csLL-PRB, castro-TBG-network-PRB} and express $\tilde{K}_{cS}$ as periodic function of $q_{\perp}$,
\begin{align}
    \tilde{K}_{cS}(q_{\perp}) =  K_{cS} \left[1+\lambda_{1}\cos(q_{\perp}d)+\lambda_{2}\cos(2q_{\perp}d)\right],
    \label{eq:tildeK_cs}
\end{align}
which we keep the first three Fourier components for simplicity.
Here we introduce the dimensionless parameters $K_{cS} = \left(1 + \frac{U_{\rm ee}}{\pi \hbar v_{\rm dw}}\right)^{-1/2}$~\cite{Giamarchi2003} and $\lambda_{1,2}$ to characterize the interaction strength, with $U_{\rm ee}$ estimated from the screened Coulomb interactions. The values of $\lambda_{1,2}$ are bounded by the condition $\tilde{K}_{cS}(q_{\perp}) > 0$ for $-\pi \le q_\perp d \le \pi $. 
The detailed derivation is shown in Supplemental Material (SM)~\cite{supp}.

Interestingly, Eq.~\eqref{eq:Hee} generalizes the coupled-wire or sliding Luttinger liquid (SLL) Hamiltonian, as previously proposed for cuprates~\cite{Emery-sLL-2000,Carpentier-sLL-2001,Mukhopadhyay-sLL-PRB-2001,Mukhopadhyay-csLL-PRB}. Alternative bosonized models for TBG have also been proposed~\cite{Cenke-PRB-moire-coupledWire,YCChou-PRB-moire-network,Gil-PRL-DW-moire-NFL}, although they do not adopt the SLL description used here. Importantly, since the interaction-to-bandwidth ratio $U_{\rm ee}/(\hbar v_{\rm dw})$ can be estimated from the continuum model~\cite{HCWang-TBG}, the parameter $K_{cS}$ can also be evaluated accordingly and exhibits electrical tunability---an advantage in two-dimensional \moire materials.
 
Before proceeding, we note that our analysis focuses on the low-temperature regime within the energy window set by the biased-induced local spectral gap in the AB- and BA-stacking regions~\cite{HCWang-TBG}. In this regime, only domain wall modes participate in scattering processes, as the bulk modes in the \moire  bands are gapped out~\cite{San-Jose-TBG-network,Efimkin-2018-TBG-network,Koshino-moireDW-2020,moire-DW-Ensslin,LeRoy-moire-DW-exp,PhilipKim-moire-DW,Xu-moireDW-GiantOscil,Carr-DW-hetero,Peeters-PRMater,Fleischmann-NanoLett-2020,Ren-PRB-Metallic-network,Walet-2020,Verbakel-PRB}.
Additionally, although scattering involving crossing domain walls may occur at their intersections, such processes are generally less relevant than those involving single or multiple parallel domain walls, since the corresponding operators enter the effective action without a spatial integral~\cite{castro-TBG-network-PRB,Hsu:2023}. We therefore focus on scattering within a single domain wall or between parallel domain walls, and omit the array index in the following.

The Kondo-type interaction $H_{{\rm K}}$ between conduction electrons and localized moments is given by 
\begin{align}
   H_{{\rm K}}=& \sum_{k,m}\sum_{\mu , \sigma , \sigma^{\prime}} \frac{J_K^{\mu} }{N_{\perp}} \left[ 
   \psi^{\dagger} _{\sigma,m}(r_{k}) \sigma^\mu_{\sigma\sigma^{\prime}}
   \psi_{\sigma^\prime,m}(r_{k}) \right]   S^\mu_{m} (r_{k}) \;,
   \label{eq:HK}
\end{align}  
with the coupling  $J_K^{\mu}$ with $\mu \in \{x,y,z\}$ (taking into account reduced $J_K^{x,y}$ for domain wall modes~\cite{supp}), the number $N_\perp$ of localized moments within the transverse extent of the domain wall modes $\psi_{\sigma,m} \equiv \sum_{\ell,\delta}\psi_{\ell \delta\sigma,m}$, and the spin $\bm{S}_{m}(r_{k})$ (modeled as classical spins) near the domain wall with length $S$ at the position labeled by $r_k$. Note that the interaction of this type in Eq.~\eqref{eq:HK} is not unique to the Kondo interaction in magnetic alloys or heavy fermions~\cite{Hewson_book}, as the hyperfine interaction also takes the same form. We also remark that the dipole-dipole interaction between the localized moments is significantly weaker than that of $H_{\rm ee}$ and $H_{\rm K}$~\cite{Paget-PRB-1977}, 
and therefore not included in our analysis.

To proceed, we focus on the weak-$J_K$ regime~\cite{Braunecker-2009,Braunecker-PRL-2009-C13,Meng-2013-PRB-2-subband,Jelena-2013-PRL-RKKY-TSC,Meng-strip-EuroPhys,CHH-CNT-helix} and perform the Schrieffer-Wolff transformation on $H_{\rm K}$ by integrating out the electron degree of freedoms. Retaining terms up to second order in $J_K$, we obtain an indirect Ruderman-Kittel-Kasuya-Yosida (RKKY) interaction between localized moments ~\cite{supp},  
\begin{align}
    H_{\rm R}= \sum_{m,n}\sum_{\mu,\nu  } & \int  
    \frac{ drdr^\prime }{N_{\perp}^{2}}
    J_{n}^{\mu }(|r-r^\prime|)
    S^{\mu}_{m+n}(r)S_m^{\mu}(r^\prime) \;,
    \label{eq:RKKY}
\end{align}
with the spin operator $ S_{n}^{\mu}(r_{k})/a \to S_{n}^{\mu}(r)$ in continuum limit, short-distance cutoff $a$, and the spatially oscillating coupling strength $J_{n}^{\mu}$ proportional to the spin susceptibility of electrons between the $n$th-nearest-neighbor parallel domain walls. 
It is justified to focus on the weak-$J_K$ regime here, since, for typical parameters, the Kondo temperature is well below all other relevant scales~\cite{Braunecker-2009,Braunecker-PRL-2009-C13,CHH-CNT-helix}, so the localized spins are governed solely by the indirect RKKY interaction $H_{\rm R}$.
In momentum space, $J_{n}^{\mu}$ develops dips at momenta corresponding to scattering processes involving single or multiple parallel domain walls.  As shown in Fig.~\ref{fig:schematic}(b), these backscattering processes include the intrabranch ($\intra$) processes with momentum transfer (projected onto the domain wall) $q = \pm 2k_{F_{1,2}} $ and the interbranch ($\inter$) ones with $q= \pm  2Q_F \equiv \pm (k_{F_1} + k_{F_2})  $. 
The $\pm 2Q_F$ processes develop a global maximum (in absolute value) for $q_\perp = 0$, owing to more available states for scatterings~\cite{supp}. 
To minimize energy, the localized moments tend to align with the Fourier component corresponding to the dip position, $q  =  \pm 2Q_F$. Below we discuss the ordering of these localized spins.

\textit{Spin helix formation and magnon spectrum.}
We now demonstrate that the localized moments tend to form a helical pattern with spatial period $\pi/Q_F$ along the domain walls.  Given the inherent $C_3$ rotational symmetry of our model, there is no preferred direction for the formation of the quasi-2D spin helix, making it equally probable to develop in any of the three arrays. Formally, we take the ansatz incorporating an offset phase $\vartheta_m$ depending on the domain wall index,
\begin{align}
    & \langle \bm{S}_m (r)\rangle =
     m_{2Q_F} (T)  S N_\perp / a  \nn 
    & \hspace{26pt} \times \left[\hat{x}\cos\left(2Q_{F}r +\vartheta_m\right)  
    + \hat{y}\sin\left(2Q_{F}r+\vartheta_m\right)\right]\;,
    \label{Eq:helical-ansatz}
\end{align}
with the order parameter $m_{2Q_F}$ satisfying $m_{2Q_F}(0)=1$ and $m_{2Q_F}(T_{\rm hx})=0$, and the ordering temperature $T_{\rm hx}$. 
Before proceeding, we discuss the spatial rotational symmetry breaking in the ansatz. Namely, simultaneously establishing a helix in all three arrays would require the helix period $\pi/Q_F$ to be commensurate with $\lambda_M$ [see Fig.~\ref{fig:schematic}(a)], as the triangular network would otherwise lead to geometrical frustration.
However, this commensurate condition necessitates precise tuning of the chemical potential. Under typical conditions, a quasi-2D spin helix will form within a single array as shown in Fig.~\ref{fig:schematic}(d), thereby breaking the $C_3$ rotational symmetry---a scenario we explore throughout the article.  

To proceed, we derive the magnon spectrum from Eqs.~\eqref{eq:RKKY}--\eqref{Eq:helical-ansatz} using spin-wave analysis, retaining only the leading-order terms in the small parameter $1/(N_{\perp}S) \ll 1$. This procedure leads to a 2-by-2 matrix, whose twice-positive eigenvalue gives the magnon dispersion~\cite{supp}, 
\begin{align}
   & \hbar  \omega(q,q_{\perp})  =  \frac{S}{2^{3/2}  N_{\perp}} \sqrt{ J^x_{q_\perp = 0} (2Q_F) - J^z_{q_\perp} (q)}  \nn
   & \times \sqrt{2J^x_{q_\perp = 0} (2Q_F) - J_{q_\perp}^x (2Q_F +q) -J_{q_\perp}^x (2Q_F -q) } ,
   \label{eq:magnon-dispersion}
\end{align}
where $J^\mu_{q_\perp}(q)$ is the Fourier transform of $J^\mu_n(r)$. The resulting magnon spectrum for representative parameters is shown in Fig.~S4 of the SM~\cite{supp}.
For $q_{\perp} =0$, Goldstone zero modes are present at  $q= 0, \; \pm 2 Q_F $, corresponding to the breaking of spin rotational symmetry in the helix phase~\cite{Jelena-2013-PRL-RKKY-TSC,CHH-CNT-helix}; for $q_{\perp}  \neq 0$, the Goldstone modes acquire finite energy.
In realistic systems, however, these zero modes can be gapped due to finite-size effects~\cite{Meng-strip-EuroPhys,2018-CHH-nuclear-2DTI}, circumventing the Mermin-Wagner theorem for the thermodynamic limit.
Namely, in a domain wall of length $L$, the momentum is quantized in unit of $q = \pi/L$ and the magnon energy is approximately constant (see Fig.~S4 in SM~\cite{supp}),  $\hbar\omega (q=\frac{\pi}{L},q_\perp)  \approx 
\hbar\omega_{0} \equiv S\left|J^x_{q_\perp = 0} (2Q_{F})\right| / 2N_{\perp}  $, set by the RKKY energy scale $J^x_{q_\perp =0}(2Q_F)$.

Next, we obtain the magnetic energy gain from Eqs.~\eqref{eq:RKKY}--\eqref{Eq:helical-ansatz}, highlighting a key difference compared to isolated channels~\cite{Meng-strip-EuroPhys,2018-CHH-nuclear-2DTI}.
Specifically, owing to the nonlocal contributions $J_{n \neq 0}^{\mu}$ in Eq.~(\ref{eq:RKKY}), the energy is minimized when the offset $\vartheta_m$ is uniform across domain walls~\cite{supp}. 
Combined with the fact that the global maximum of $|J_{n \neq 0}^{\mu}|$ occurs at $q_\perp = 0$, this leads to the development of spatial phase coherence among spin helices in distinct domain walls in parallel to each other; see Fig.~\ref{fig:schematic}(d). 
With the numerically computed magnon energy $\hbar\omega_{0} $, we estimate the ordering temperature $T_{\rm hx}$ while self-consistently incorporating effects of the spatially rotating magnetic field induced by spin ordering~\cite{supp}. Notably, reflecting the 2D nature of the system, the contributions from $J_{n \neq 0}^{\mu}$ result in an increase in $T_{\rm hx}$ more than an order of magnitude. 
Additionally, the helix-induced field couples to the electron spins, opening a partial gap in the domain wall spectrum. This gap leads to a Peierls energy gain, further stabilizing the spin helix.
Consequently, we establish that the system forms a $C_3$-breaking spatially phase-coherent quasi-2D spin helix at sufficiently low temperatures.

\begin{figure}[t]
    \centering     \includegraphics[width=0.48\textwidth]{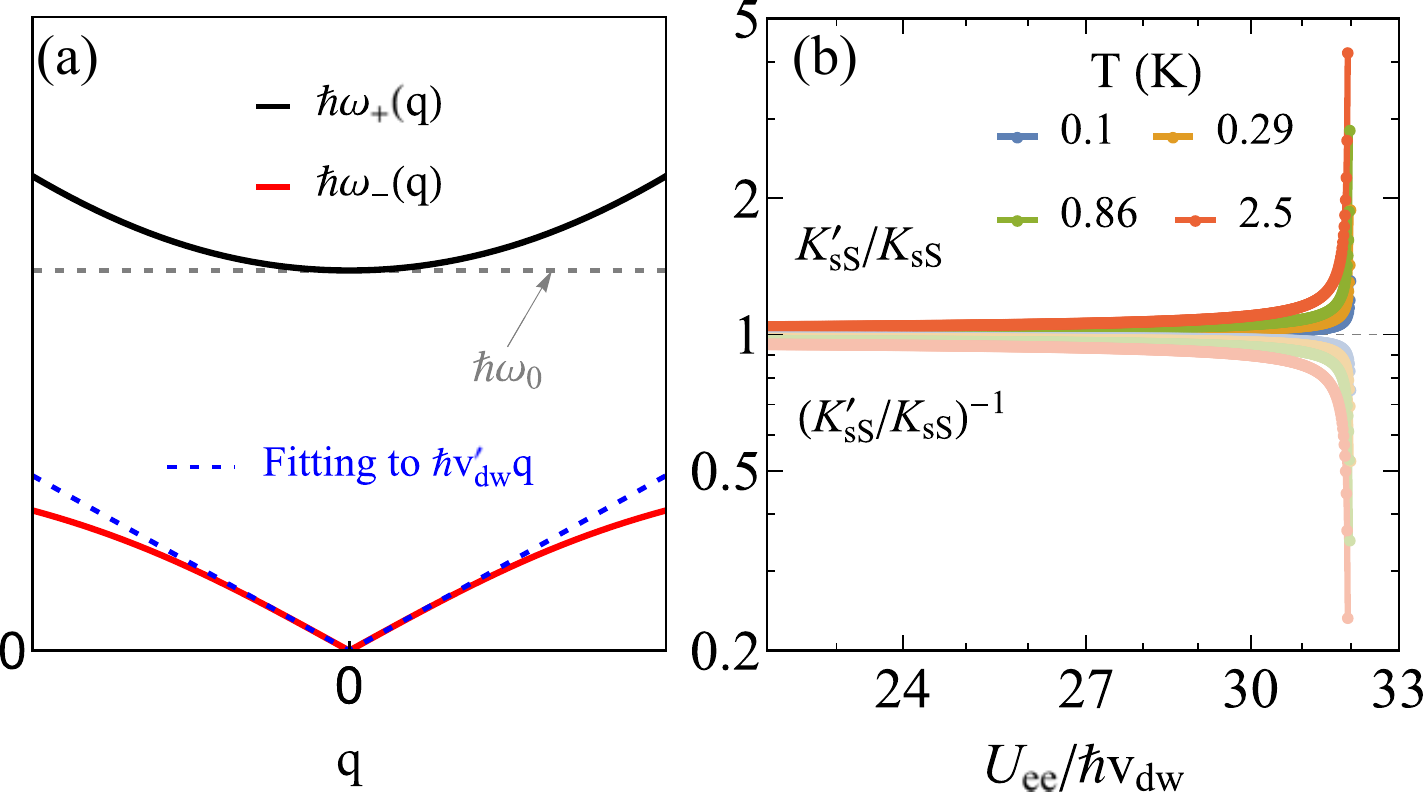}
    \caption{(a) Excitation spectrum $\hbar \omega_\pm (q)$ (black/red curves) in electron-magnon-coupled systems. For small $q$, the lower band $\omega_- (q)$ follows linear dispersion with velocity $v_{\rm{dw}}^\prime$ (blue dashed line). (b) Modified scaling exponent ($K^\prime_{sS}$) as a function of the   interaction strength $ (U_{{\rm ee}}) $ for various temperatures and $L = 0.5 \, \mu \rm {m}$, $N_\perp = 80$, $J_K = 1 \; $meV, $\lambda_1 = \lambda_2 = 0.2$, and $N_{\rm dw} = 20$. 
    }
    \label{fig:singularity-v2}
\end{figure}

\textit{Magnon-induced singularity.}
In the spin helix phase, the magnons can lead to spin flips of electrons, which is known to influence electrical transport through backscattering~\cite{Nuclear-localization-2017-PRB,2018-CHH-nuclear-2DTI,Hsu:2021}. Here, we instead look into the magnon-induced forward scattering and explore its effects on the scaling dimensions of various operators. 
To this end, we express Eq.~(\ref{eq:HK}) as  $H_{\rm K} \approx  \left\langle H_{{\rm K}}\right\rangle_{\rm hx} +H_{{\rm em}}$, with the expectation value $ \left\langle \cdots \right\rangle_{\rm hx}$ with respect to the spin helix phase. The coupling between the electron spin density and the magnon-induced spin flip can be formulated as 
\begin{align}
   H_{\rm em} = g_{{\rm em}}\int\frac{dr}{2\pi}\sum_{m}\left[\partial_{r}\phi_{sS,m}(r)\right]\Pi_{m}(r)\;,
   \label{eq:Hem}
\end{align}
with the electron-magnon coupling strength $g_{{\rm em}}\equiv-2J_{K}\sqrt{a S m_{2Q_F}/(\hbar\omega_{0}N_{\perp})}$, $\Pi_{m}(r)\sim [a_{m}(r)-a_{m}^{\dagger}(r) ]$, and magnon field $a_{m}$. 
In the bosonic language, magnons couple to the spin-symmetric boson  ($\phi_{sS,m}$) within each domain wall, which can therefore influence the electron subsystem.  
We present the corresponding excitation spectrum in Fig.~\ref{fig:singularity-v2}(a), which is characterized by a gapped $\omega_{+}$ branch with band bottom $\omega_0$ and a gapless branch $\omega_{-}$ with modified velocity $v_{\rm dw}'$.

Since the electron-magnon coupling enters the Hamiltonian in quadratic form, we diagonalize the full system and compute the scaling dimension $K_{sS}^\prime \, \left(1/K_{sS}^\prime \right)$ of the operator $e^{i \phi_{sS}} \, \left(e^{i \theta_{sS}}\right)$. This allows us to compare them to the one  without magnons,
 \begin{align}
    \frac{K^\prime_{sS}}{K_{sS}}
    =\left(1-\frac{J_K^2 a SK_{sS}m_{2Q_F}}{\hbar^2 v_{\rm dw}\omega_0}\right) ^{-\frac{1}{2}}\;.
    \label{Eq:KsS-modification}
\end{align}
As illustrated in Fig.~\ref{fig:singularity-v2}(b), 
a singularity arises when the quantity in the parenthesis of Eq.~\eqref{Eq:KsS-modification} vanishes. This is reflected in the divergence (vanishing) of the modified scaling dimension of the correlation function of $\phi_{sS}\,(\theta_{sS})$. The magnon-induced singularity can be observed through physical quantities such as paramagnetic susceptibility or spin relaxation rate, which depends on the renormalized parameter ${K}_{sS}^\prime$ and carrier velocity $v_{\rm dw}^\prime \propto 1/K^\prime_{sS}$, and will be discussed later. The singularity can be experimentally accessed by adjusting the temperature and the interaction strength, $U_{\mathrm{ee}}$, the latter of which is tunable through twist angle and interlayer bias~\cite{HCWang-TBG}. More precisely, increasing the distance from a metallic gate, decreasing the twist angle, and/or increasing the bias voltage enhances the ratio $U_{\rm ee}/(\hbar v_{\rm dw})$; see SM~\cite{supp} for details. 
In contrast, we found that the exponents are robust against $J_K$ (as $\omega_0 \propto J_K^2$) and the interaction parameters $\lambda_{1,2}$. A similar divergence driven by phonons has been discussed in (quasi-)one-dimensional systems~\cite{Bardeen-RMP-WB,Wentzel-PR-WB,Loss-WB,Martin-WB-PRB,CHH-nano-horizon-2024, HCWang-TBG}.

\textit{Realization and transport features.}
Having demonstrated the general picture of the spin helix formation, we now discuss two scenarios for its realizations. The first one is TBG fabricated using $^{13}$C isotopes, in which conduction electrons and nuclear spins couple through the hyperfine interaction.
While detailed investigations of hyperfine coupling, similar to studies on semiconductors~\cite{Schliemann2003}, remain absent for \moire systems, it is noteworthy that an experimental hyperfine coupling strength of $O(100~\mu$eV) has been reported in nanotubes~\cite{Churchill2009-naturephysics}, exceeding the theoretical value~\cite{Fischer2009}. 
We estimate $T_{\rm hx} = O(10~\mathrm{mK})$ for $J_K = O(\mu\rm{eV})$  and typical parameters for the electron subsystem (i.e., those adopted for Fig.~S3 in SM~\cite{supp}). 
The second scenario involves magnetic adatoms deposited on graphene layers, interacting with domain wall modes through exchange coupling. This has some parallels with previous studies on monolayer graphene~\cite{Castro-PRL-graphene-Kondo,Vojta-RPP-graphene-Kondo,Rosch-PRB-Co-graphene-Kondo,Co-graphen-Nano-Lett,F-adatom-graphene}, where an exchange coupling of $5$~meV has been observed in samples with fluorine adatoms~\cite{F-adatom-graphene,Vojta-RPP-graphene-Kondo}.
In this scenario, the exchange coupling generally exceeds the hyperfine coupling considered in the first scenario, and it can be further enhanced due to the increased electron density within the domain walls. We find that $T_{\rm hx}$ can reach up to $O(\rm{K})$ for $J_K = O(\rm{meV})$.

While the mesoscopic length scales discussed here should enable spin-sensitive scanning probes~\cite{2012Togawa-LorentzMicroscope,
Dussaux2016,Yu2010} to image helix formation, the presence of metallic gates may render this approach inapplicable. 
We therefore search for additional observable features. 
Since the spin helix can generate a spatially rotating magnetic field, which gaps out half of the electron modes, we expect the quantized conductance, for instance in setups in Ref.~\cite{ZheHou-2024-moire}, to reduce with onset at $ T_{\rm hx}$, providing an indirect probe for the spin helix formation~\cite{Braunecker-2009,Braunecker-PRL-2009-C13, CHH-CNT-helix,Hsu:2020}. 
Similar conductance reduction has been observed in GaAs quantum wires~\cite{Zumbuhl-2014-PRL}. 
Moreover, one can probe the spin helix and the magnon-induced singularity through magnetic properties such as magnetic resonance and paramagnetic susceptibility, which we discuss next.

\begin{figure}[t]
    \centering
    \includegraphics[width=0.43\textwidth]{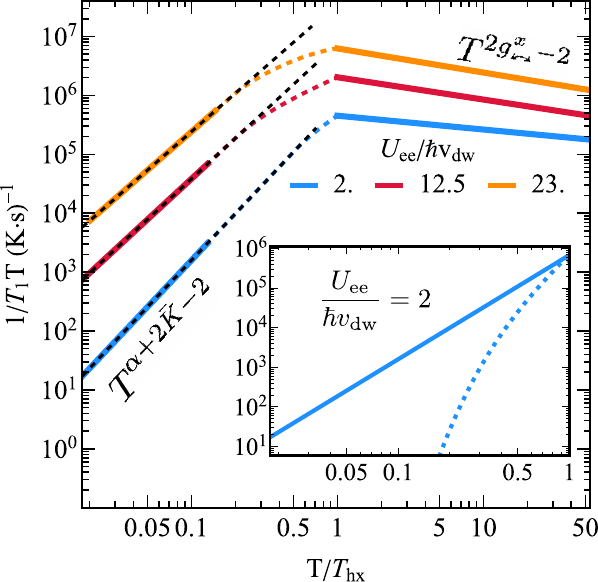}
    \caption{ Temperature ($T$) dependence of spin relaxation rate $1/(TT_1)$ for various interaction strengths ($U_{\rm ee}$). The ordering temperature $(T_{\rm hx})$ separates two regimes described by Eq.~\eqref{eq:pwr-spinrelaxT-1}--\eqref{eq:pwr-spinrelaxT-2}. The black dashed lines represent the power-law fit for the low-$T$ regime, while the colored dashed curves serve as visual guides for $T \lesssim T_{\rm hx}$. The inset shows the contributions from gapped (dashed) and remaining gapless (solid) modes. The other parameter values are given in the caption of Fig.~\ref{fig:singularity-v2}. 
    }
    \label{fig:spinrelaT}
\end{figure}

\textit{Magnetic resonance.}
The transition into the quasi-2D spin helix state, along with the properties on both sides of the transition, can be detected by measuring the spin relaxation rate, $1/T_1$, in magnetic resonance experiments\footnote{The spin relaxation rate investigated here also supplements the predicted absorption frequency in Ref.~\cite{Stano-NMR-absorption}, which is governed by the RKKY energy scale with external magnetic fields producing side peaks. 
}.
This rate captures the local dynamics of the magnetic moments due to the exchange interaction $H_{\rm K}$ and displays two regimes as displayed in Fig.~\ref{fig:spinrelaT} and described by 
\begin{subnumcases}
{\frac{1}{T T_1 } \propto}
T^{2g_{{\inter}}^{x}-2}, & $T>T_{{\rm hx}}$, 
\label{eq:pwr-spinrelaxT-1} \\
\left[ 1 - m_{2Q_F}(T) \right] T^{2\bar{K}-2} ,  
 & $T \ll T_{{\rm hx}},$ 
 \label{eq:pwr-spinrelaxT-2} 
\end{subnumcases}
where $(1-m_{2Q_F}) \propto T^\alpha$ with a numerical  exponent $\alpha$ obtained from fitting. Here, $g_{{\inter}}^{x}$ and $\bar{K}$ are the fractional power-law exponents of the spin susceptibility for temperatures above and below $T_{\rm hx}$, respectively. Their explicit forms are determined by the interacting strength through  $K_{cS}$   (see SM~\cite{supp}):  
\begin{subequations}
    \label{eq:g-and-Kbar}
\begin{eqnarray}
    g_{{\inter}}^{x}&=&\frac{1}{4}\left( \bar{\Delta}_{\phi_{cS},n=0}+3\right) \;,\\
    \bar{K} & = & \int^\pi_{ {\pi}/{N_{\rm dw} }} \frac{d(q_\perp d)}{\pi} \sqrt{\frac{\tilde{K}^2_{cS}(q_\perp)+3\tilde{K}_{cS}(q_\perp)}{3\tilde{K}_{cS}(q_\perp)+1}}\;,
    \end{eqnarray}
\end{subequations}
where $\bar{\Delta}_{\phi_{cS},n}\equiv (2
\pi)^{-1} \int^\pi_{-\pi} d(q_\perp d) \tilde{K}_{cS}(q_\perp)\cos(nq_\perp d)$. In Eq.~\eqref{eq:g-and-Kbar}, the condition $K_{\nu P \neq cS} = 1$ has been used. We note that the analysis here can also be generalized for  isolated channels. 
Above $T_{\rm hx}$, the SLL phase exhibits a power-law temperature dependence in Eq.~\eqref{eq:pwr-spinrelaxT-1}, characterized by a fractional exponent $g^x_{\inter}$, generalizing the conventional Korringa law for Fermi liquids~\cite{Korringa1950,Coleman_book}, similarly to Rashba nanowires~\cite{meng-nmr-rashbar-wire}.

\begin{table*}[t]
\caption{Paramagnetic spin susceptibility $\chi_u$ in terms of the overall scale, $  \chi_u^0 \equiv \frac{\mu_{0}\left(g\mu_{B}\right)^{2}}{\pi\hbar v_{{\rm dw}} d} $. Here, we have $\tilde{K}_{cS}(q_\perp = 0) = K_{cS}(1+\lambda_1 + \lambda_2)$ from Eq. \eqref{eq:tildeK_cs}.
The detailed derivation is provided in Sec. S.VI of SM~\cite{supp}.
    }
    \centering
    \setlength{\tabcolsep}{6pt} 
    \renewcommand{\arraystretch}{1.7} 
    \begin{tabular}{c c c c}
        \hline\hline
         Temperature regime    & $T \to 0 $ & $T\ll  T_{\rm hx} $ & $T >  T_{\rm hx} $  \\ \hline
        \addlinespace[8pt]
       Susceptibility $\chi_u/\chi_u^0$    
       & $\displaystyle \frac{K_{sS}\cos^2 \varphi}{1 +\tilde{K}_{cS}(q_\perp = 0)K_{sS}} $  & $\displaystyle \frac{K_{sS}\cos^2 \varphi}{(K_{sS}/K_{sS}^\prime)^2 +\tilde{K}_{cS}(q_\perp = 0)K_{sS}} $ & $\displaystyle K_{sS}$    \\[0.5cm]
       \hline \hline
    \end{tabular}
    \label{tab:Thx}
\end{table*}

Sufficiently below $T_{\rm hx}$, the decay takes a distinct form shown in Eq.~(\ref{eq:pwr-spinrelaxT-2}), where the factor $(1-m_{2Q_F})$ weights the disordered part of the localized spins. 
As mentioned, the formation of a helix induces a partial gap in the domain wall spectrum, which suppresses the relaxation channel from the gapped modes, giving rise to a rate of exponential form. As shown in the inset of Fig.~\ref{fig:spinrelaT}, the relaxation is thus dominated by the remaining gapless modes, characterized by the effective parameter $\bar{K}$.
Consequently, we obtain a generalized power-law decay distinct from the conventional exponential suppression in fully gapped systems~\cite{Single-CNT-NMR-PRL,Dora-PRL-NMR-CNT-spinrelaxation}.
Additionally, upon approaching the magnon-induced singularity, the excitation velocity vanishes and the density of states becomes singular, leading to a divergence in the low-temperature regime of Fig.~\ref{fig:spinrelaT}.

For temperatures slightly below $T_{\rm hx}$, the crossover behavior in this regime (dashed curves in Fig.~\ref{fig:spinrelaT}) may require further analysis beyond our approach~\cite{Vayrynen-MagneticHelicalEdge-PRB-2016,Hsu:2021}. 
To access this additional crossover scale, one should consider the dynamics of the local moments and the increased thermal population of magnons, which may influence the polarization of the local spins and thereby affect the spin relaxation rate.
Nevertheless, the primary prediction here is the distinct power laws in Eqs.~\eqref{eq:pwr-spinrelaxT-1}--\eqref{eq:pwr-spinrelaxT-2}.
Notably, our numerical results for typical parameters indicate that the power law in the helix phase is generally steeper (specifically, $\alpha + 2\bar{K} - 2 > 2 - 2g^x_{\inter}$). 
Given the nanoscale nature of our target systems,  the predicted features can be detected through resistively-detected spin resonance experiments~\cite{Resistive-NMR-Klitzing,Book-Resistive-NMR,muraki-science-2006,muraki-2007-PRL,muraki:2012,muraki:2014,RDESR-TBG}.

\textit{Anisotropic paramagnetic susceptibility.}
We examine the paramagnetic susceptibility $\chi_{u} = \mu_0  (\partial M/\partial B)$ with the permeability $\mu_0$, the total magnetization $M$ of the electron subsystem, and  the in-plane external magnetic field $\bm{B}=B\hat{\bm{n}}$ forming an angle $\varphi$ with the domain wall direction [where $\hat{\bm{n}}=(\sin \varphi, \; 0, \; \cos\varphi)$]. 
Taking into account the helix-induced effective field and electron-magnon coupling, we derive the contribution to the paramagnetic susceptibility, as summarized in Table \ref{tab:Thx} and detailed in Sec. S.VI. of SM~\cite{supp}. 
In particular, for $T < T_{\rm hx}$ we get a $\cos^2 \varphi$ dependence of the anisotropic susceptibility, indicating the spontaneous breaking of rotational symmetry. 
The maximum $\chi_u$ thus identifies the specific domain wall direction where the spin helix forms, with magnitude depending on the $e$-$e$ interaction and the \moire pattern. Notably, as the magnon-induced singularity is approached [see Fig.~\ref{fig:singularity-v2}(b)], the paramagnetic susceptibility peaks, serving as an experimental indicator of the magnon-induced singularity.

 
\textit{Conclusions.}
We demonstrate the formation of a spatially phase-coherent planar spin helix and a magnon-induced singularity in TBG networks with observable features. 
Owing to its 2D nature, direct probing of the helix formation or magnon spectrum with inelastic neutron scatterings~\cite{lovesey-neutron-1984-theory,Helical-spin-theory-NeutronDiff,DyY-ND-PRL,Squires-Neutron-2012,willis-2017-neutron-exp,Furrer-neutron}, resonant inelastic X-ray scatterings~\cite{Luuk-RMP-RIXs,RMP-dynamic-XRS,Brink-RIXS,Fatale-RIXS-review} or Lorentz microscope~\cite{2012Togawa-LorentzMicroscope} might not be practical.
Alternatively, spin resonance and paramagnetic susceptibility provide viable options; since the former is extensive, fabricating sizable TBG samples will be advantageous~\cite{NCYeh-TBG-growth}. Given that conventional 2D Kondo lattice models typically describe noninteracting electrons, observing the features predicted here could unveil an electrically tunable platform where interacting electrons and  Kondo physics interplay. This approach is instrumental in revealing various quantum phases, including skyrmion lattices stabilized by external magnetic fields~\cite{Okubo-PRL-2012,Togawa2016},
strange metals~\cite{Komijani2018,Komijani2019,2020-JF-PRB,sm-phase-PNAS} and topological heavy fermion superconductivity~\cite{Hazra-UTe2}.

\let\temp\addcontentsline 
\renewcommand{\addcontentsline}[3]{} 

{\it Data availability statement.} 
The data that support the findings of this article are openly available~\cite{data}.

 
We acknowledge interesting discussions with G.~Boebinger, G.~Bihlmayer, C.-K.~Chang, C.-D.~Chen, S.-Y.~Chen, Y.~Kato, C.-T.~Ke, D.-J.~Huang, Y.~Togawa, and K.~Totsuka. We thank H.-C.~Wang for creating the symbol ($\intra$) for the intrabranch contributions. We acknowledge support from the National Science and Technology Council (NSTC), Taiwan through Grant No.~NSTC-112-2112-M-001-025-MY3, Grant No.~NSTC 112-2811-M-001-061 and Grant No.~NSTC-114-2112-M-001-057, and Academia Sinica (AS), Taiwan through Grant No.~AS-iMATE-114-12. C.-H.H. acknowledges support from the National Center for Theoretical Sciences (NCTS), Taiwan. Y.-Y.C. acknowledges the financial support from The 2023 Postdoctoral Scholar Program of Academia Sinica, Taiwan. K.S. acknowledges the financial support from JST SPRING (Grant No.~JPMJSP2151).



\putbib[DW-TBG-bib]
\let\addcontentsline\temp 
\end{bibunit}


\clearpage
\onecolumngrid
\bigskip

\begin{bibunit}[apsrev4-2]

\setcounter{section}{0}
\setcounter{subsection}{0}
\setcounter{subsubsection}{0}
\setcounter{equation}{0}
\setcounter{figure}{0}
\setcounter{table}{0}
\setcounter{page}{1}
\setcounter{NAT@ctr}{0} 
\makeatletter
\renewcommand{\thesection}{S.\Roman{section}}
\renewcommand{\theequation}{S\arabic{equation}}
\renewcommand{\thefigure}{S\arabic{figure}}
\renewcommand{\bibnumfmt}[1]{[S#1]}
\renewcommand{\citenumfont}[1]{S#1}

\begin{center}
\large{\bf Supplemental materials for  ``Quasi-two-dimensional spin helix and magnon-induced singularity in twisted bilayer graphene''}\\
\vspace{15pt}
\fontsize{10}{12}
Yung-Yeh Chang$^{1}$, Kazuma Saito$^{1,2}$ and Chen-Hsuan Hsu$^{1}$\\
\textit{$^{1}$Institute of Physics, Academia Sinica, Taipei 11529, Taiwan}\\
\textit{$^{2}$Department of Applied Physics, Tokyo University of Science, Katsushika, Tokyo 125-8585, Japan}
\end{center}

\tableofcontents

\resumetoc

\section{Sliding Luttinger liquid description of the domain wall network on TBG}
\label{supp:H_ee}

In this section, we describe the electron subsystem of our model, which consists of a domain wall network which is populated by interacting electrons~\cite{HCWang-TBG}.
In the fermionic expression, we have 
\begin{align}
    H_{\mathrm{ee}} &= \sum_{jm, \ell\delta  \sigma} \int \dd{r} [\psi^{(j)}_{\ell\delta\sigma, m}(r)]^\dagger (-i\hbar v_{\mathrm{dw}} \ell \partial_r) \psi^{(j)}_{\ell \delta\sigma, m}(r)
   +\sum_{\mathfrak{M}, m, n} \int \dd{r} \int \dd{r'} U_{n} (r-r ') \rho_{m}^{\mathrm{dw}}({r}) \rho_{m + n}^{\mathfrak{M}}( {r}'), 
    \label{Eq:fermion}
\end{align}
with the fermion field
$\psi_{\ell \delta \sigma, m}^{(j)}$ with the indices given in Fig.~1(b) caption in the main text. The first term represents the kinetic energy of electrons with the velocity $v_{\mathrm{dw}}$. The second term is the screened Coulomb potential $U_{n}( r )$ between the $n$th nearest neighbor domain walls as illustrated in Fig.~1(c) in the main text, and the electron density $\rho_{m}^{\mathfrak{M}}$, where $\mathfrak{M} \in \{\text{dw, image}\}$ refers to contributions from domain wall electrons and their image charges.

To proceed, we describe the electrons along a domain wall labeled by $m$ in the $j$th array in the following bosonized form,
\begin{align}
    \psi_{\ell\delta\sigma,m}^{(j)}(r)=\frac{U_{_{\ell\delta\sigma,m}}^{j}}{\sqrt{2\pi a}}e^{i\ell k_{F_{\delta},m}^{(j)}r} & \exp\left\{ \frac{i}{2} \left[-\ell\left(\phi_{cS,m}^{j}+\delta\phi_{cA,m}^{j}\right)-\ell\sigma\left(\phi_{sS,m}^{j}+\delta\phi_{sA,m}^{j}\right)\right.\right.\nn
    & \left.\left.\qquad  \qquad \qquad   +\left(\theta_{cS,m}^{j}+\delta\theta_{cA,m}^{j}\right)+\sigma\left(\theta_{sS,m}^{j}+\delta\theta_{sA,m}^{j}\right)\right]\right\} ,
\label{eq:bosonized-fermion}
\end{align}
where $a$  denotes the short-distance cutoff,  the symmetry index $P$ is defined as  $P\in \big\{ S \,:{\rm symmetry}, \; A\, : {\rm antisymmetry};\,S=+1,\,A=-1 \big\}$ and $U_{_{\ell\delta\sigma,m}}^{j}$ represents the Klein factor.  
The electron subsystem is described by $H_{\rm{ee}}$, composed of a kinetic energy term $H_{0}$ and interaction terms   within domain walls $(V_{\parallel})$ and  across parallel domain walls $(V_{\perp})$,
\begin{align}
    H_{\rm{ee}} = \sum_j H_{\rm{ee}} ^{(j)} = \sum_j \left[  H_{0} ^{(j)} +V_{\parallel} ^{(j)}  +V_{\perp} ^{(j)} \right] \;.
    \label{eq:H_ee-simple}
\end{align}
The interaction term across different arrays is less relevant in the renormalization-group sense~\cite{Hsu:2023} and hence not included here.

In this work, we assume periodic boundary condition perpendicular to the domain walls within each array. The three terms of Eq. (\ref{eq:H_ee-simple}) are diagonal in $\phi_{\nu P,q_{\perp}}^{j}$ and $\theta_{\nu P,q_{\perp}}^{j}$, and can be expressed as
    \begin{align}
         H_{0} ^{(j)} 	& =\frac{\hbar v_{{\rm dw}}}{N_{\rm dw}}\sum_{q_{\perp}}\sum_{\nu,P}\int\frac{dr}{2\pi}\,\left[\left|\partial_{r}\phi_{\nu P,q_{\perp}}^{j}\right|^{2}+\left|\partial_{r}\theta_{\nu P,q_{\perp}}^{j}\right|^{2}\right] \;, \nn
V_{\parallel}  ^{(j)} 	& =\frac{\hbar v_{{\rm dw}}}{N_{\rm dw}}\sum_{q_{\perp}}\sum_{\nu,P}\int\frac{dr}{2\pi}\,\left[V_{\parallel\phi_{\nu P}}^{(j)}\left|\partial_{r}\phi_{\nu P,q_{\perp}}^{j}\right|^{2}+V_{\parallel\theta_{\nu P}}^{(j)}\left|\partial_{r}\theta_{\nu P,q_{\perp}}^{j}\right|^{2}\right] \;, \nn
V_{\perp}  ^{(j)} 	& =\frac{\hbar v_{{\rm dw}}}{N_{\rm dw}}\sum_{q_{\perp}}\sum_{\nu,P}\int\frac{dr}{2\pi}\,\left[V_{\perp\phi_{\nu P}}^{(j)}(q_{\perp})\left|\partial_{r}\phi_{\nu P,q_{\perp}}^{j}\right|^{2}+V_{\perp\theta_{\nu P}}^{(j)}(q_{\perp})\left|\partial_{r}\theta_{\nu P,q_{\perp}}^{j}\right|^{2}\right] \; ,
\label{eq:H_ee-3}
    \end{align}
or, combined into 
\begin{align}
    H_{\rm ee} = \frac{1}{N_{{\rm dw}}} \sum_{j} \sum_{q_{\perp}}\sum_{\nu,P}\int\frac{\hbar dr}{2\pi}\left[\frac{v^{(j)}_{\nu P}(q_{\perp})}{\tilde{K}^{(j)}_{\nu P}(q_{\perp})}\left|\partial_{r}\phi_{\nu P,q_{\perp}}^{j}(r)\right|^{2}+v^{(j)}_{\nu P}(q_{\perp})\tilde{K}^{(j)}_{\nu P}(q_{\perp})\left|\partial_{r}\theta_{\nu P,q_{\perp}}^{j}(r)\right|^{2}\right]\;.
    \label{eq:3-Hee}
\end{align}
This is essentially a generalization of sliding (Tomonaga-)Luttinger liquid (SLL) introduced for high-$T_c$ cuprates~\cite{Emery-sLL-2000,Carpentier-sLL-2001,Mukhopadhyay-sLL-PRB-2001,Mukhopadhyay-csLL-PRB}. 
In the above, we have generalized SLL parameters $\tilde{K}_{\nu P}^{j}(q_{\perp})$ and velocity $v_{\nu P}(q_{\perp})$,  
\begin{align}
    \mathcal{V}_{\phi_{\nu P}}^{(j)}(q_{\perp})&=1+V_{\parallel\phi_{\nu P}}^{(j)}+V_{\perp\phi_{\nu P}}^{(j)}(q_{\perp}), 
    \quad \left[\text{similarly for }\mathcal{V}_{\theta_{\nu P}}^{(j)}(q_{\perp}) \right]\;,  \nn
    \tilde{K}_{\nu P}^{(j)}(q_{\perp}) &=\sqrt{\frac{\mathcal{V}_{\theta_{\nu P}}^{(j)}(q_{\perp})}{\mathcal{V}_{\phi_{\nu P}}^{(j)}(q_{\perp})}}       \;;\quad 
    v^{(j)}_{\nu P}(q_{\perp}) =v_{{\rm dw}}\sqrt{\mathcal{V}_{\phi_{\nu P}}^{(j)}(q_{\perp})\mathcal{V}_{\theta_{\nu P}}^{(j)}(q_{\perp})} \approx v_{{\rm dw}} 
    \;, 
    \label{eq:general-def-v-K}
\end{align}
depending on the transverse momentum $q_{\perp}$. 
Since screened Coulomb interactions only enter the charge symmetric sector~\cite{HCWang-TBG},  we have $\tilde{K}_{cA} = \tilde{K}_{sS} = \tilde{K}_{sA} =1$ with the corresponding  velocity given by $v_{\rm dw}$.  In the following, we restrict to scatterings occurring within a single domain wall or between parallel domain walls, as scatterings of domain wall modes across different arrays are typically less relevant~\cite{Hsu:2023}. As a result, we suppress the array index from the discussion below.

To proceed, we express the SLL parameter $\tilde{K}_{c S}(q_\perp)$ of the interacting charge symmetric sector as 
\begin{align}
   \tilde{K}_{cS}(q_{\perp}) = K_{cS}\left[1+\lambda_{1}\cos(q_{\perp}d)+\lambda_{2}\cos(2q_{\perp}d)\right] ,
   \label{eq:tilde-KcS}
\end{align}
where $\lambda_{1,2}$ are dimensionless coefficients that characterize the inter-domain-wall coupling. Their values are chosen such that $\tilde{K}_{cS}(q_{\perp}) > 0$ holds over the entire range $-\pi \le q_\perp d \le \pi $.
 
The interaction parameter can be estimated through the relation~\cite{Giamarchi2003}:
\begin{equation}
    K_{cS} = \frac{1}{\sqrt{1 + \frac{U_{\mathrm{ee}}}{\pi \hbar v_{\mathrm{dw}}}}}.
    \label{eq:Kcs and Uee}
\end{equation} 
As demonstrated in Ref.~\cite{HCWang-TBG}, the interaction strength $U_{\mathrm{ee}}$ can be adjusted through the effective hybridization parameter, $\alpha_{\mathrm{AB}}$, determined by the interlayer hybridization and twist angle, as well as the interlayer bias $V_d$, dielectric material layer, and the distance $d$ from the closest metallic gate.
 
\begin{figure}
    \centering
    \includegraphics[width=0.45\linewidth]{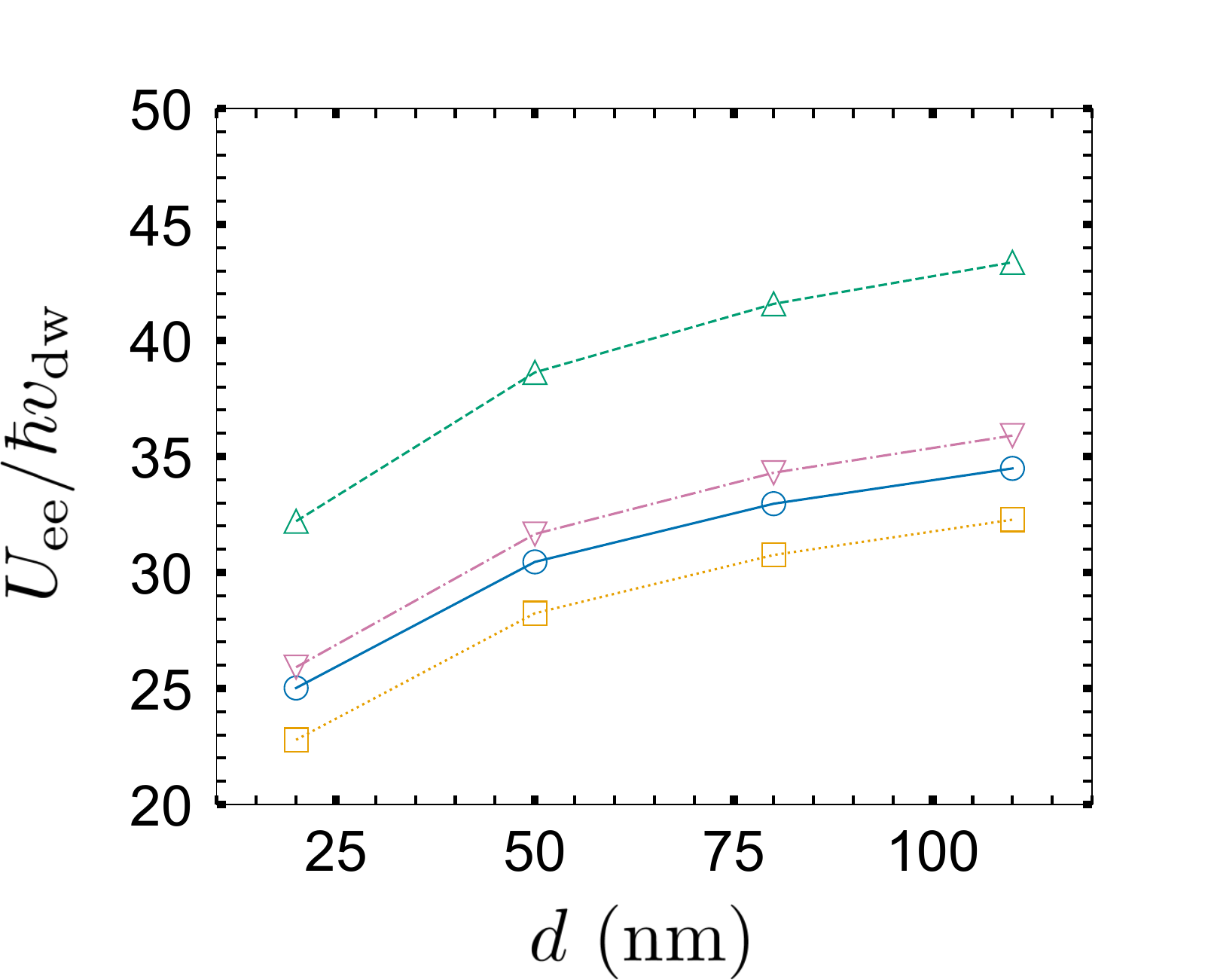}
    \caption{Interaction strength $U_{\mathrm{ee}} / \hbar v_{\mathrm{dw}}$ as a function of $d$. In our estimation, we adopt the parameter values corresponding to $(\alpha_{\mathrm{AB}}, V_{\mathrm{d}} / (\hbar v_{\mathrm{F}} k_{\theta} ))   = (1.6, 1.9)$ (blue solid), $(1.8, 1.9)$ (yellow dotted), $(1.6, 2.1)$ (green dashed), and $(1.8, 2.1)$ (purple dashed dot) following Ref.~\cite{HCWang-TBG} where we define $k_\theta = 8\pi \sin(\theta / 2) / (3a_0) $ with  the twist angle $\theta$.
    }
    \label{fig:Uee estimation}
\end{figure}

To demonstrate that the system can reach a strongly interacting regime where magnons can trigger a singularity in various correlation functions, in Fig.~\ref{fig:Uee estimation} we evaluate the interaction strength as a function of the distance $d$ between the TBG and a metallic gate for several sets of the control parameters $\alpha_{\mathrm{AB}}$ and $V_{\mathrm{d}}$.
As shown in Fig.~2(b) of the main text, the magnon-induced singularity appears when the ratio $U_{\mathrm{ee}} / \hbar v_{\mathrm{dw}}$ reaches approximately 32.
Our estimation in Fig.~\ref{fig:Uee estimation} shows that this value is well within reach through various experimentally controllable parameters.

In addition to the existence of the electrically tunable triangular network, our work differs from the SLL model in Refs.~\cite{Emery-sLL-2000,Carpentier-sLL-2001,Mukhopadhyay-sLL-PRB-2001,Mukhopadhyay-csLL-PRB} from two aspects. First, we include additional sectors (more than charge/spin), in order to reflect to the additional energy branches of the domain wall spectrum~\cite{Efimkin-2018-TBG-network, HCWang-TBG}. Second, we adopt a convention for the SLL parameter~\cite{Giamarchi2003} such that $K_{cS}<1$ indicates repulsive electron-electron interaction. 

The finite-temperature boson correlation function between $n$th-nearest-neighbor domain walls can be computed as 
\begin{align}
    \left\langle e^{-i\phi_{\nu P,m+n}(r,\tau)}e^{i\phi_{\nu P,m}(0,0) } \right\rangle_{\rm ee} = \tilde{\Omega}_{\phi_{\nu P},n}\left(\frac{a}{L}\right)^{\bar{\xi}_{\phi_{\nu P},n}/2}\frac{\left(\frac{\pi a k_B T}{\hbar v_{{\rm dw}}}\right)^{\bar{\Delta}_{\phi_{\nu P},n}/2}}{\left[\sinh^{2}\left(\frac{\pi  k_B T  r}{\hbar v_{{\rm dw}}}\right)+\sin^{2}\left(\frac{\pi k_B T  \tau}{\hbar}\right)\right]^{\bar{\Delta}_{\phi_{\nu P},n}/4}}\;,
    \label{eq:boson-correlator}
\end{align}
with the imaginary time $\tau$ and $\langle\cdots \rangle_{\rm ee}$ denoting the ensemble average with respect to the electron subsystem.
Here, the dimensionless parameters are given by 
 \begin{subequations}
 \label{eq:exponents}
\begin{eqnarray}
\bar{\Delta}_{\phi_{\nu P},n}& \equiv & \int^\pi_{-\pi}\frac{d(q_\perp d)}{2\pi}\tilde{K}_{\nu P}(q_{\perp})\cos(nq_{\perp}d)\;, \label{eq:exponents-1} \\
\tilde{\Omega}_{\phi_{\nu P},n} &\equiv& \exp\left[\frac{\bar{\xi}_{\phi_{\nu P},n}}{2}\left(\gamma-\int_{1/L}^{\infty}\frac{2\,dq}{q\left(e^{\hbar v_{{\rm dw}}q/k_B T}-1\right)}\right)\right]\;, \label{eq:exponents-2}\\
\bar{\xi}_{\phi_{\nu P},n}& \equiv& \int^\pi_{-\pi} \frac{d(q_\perp d)}{2\pi }\tilde{K}_{\nu P}(q_{\perp})\left[1-\cos(nq_{\perp}d)\right]=\bar{\Delta}_{\phi_{\nu P},n=0}-\bar{\Delta}_{\phi_{\nu P},n}\;  \label{eq:exponents-3} \;,
\end{eqnarray}
\end{subequations}
with the Euler-Maclaurin number, $\gamma \approx 0.577 $. The correlator for the $\theta_{\nu P}$ field has the same form as  Eq. (\ref{eq:boson-correlator}), but with different parameters (with $\phi \to \theta$) given by
\begin{subequations}
 \label{eq:exponents-theta}
\begin{eqnarray}
\bar{\Delta}_{\theta_{\nu P},n}& \equiv  & \int^\pi_{-\pi}\frac{d(q_\perp d)}{2\pi}\frac{1}{\tilde{K}_{\nu P}(q_{\perp})}\cos(nq_{\perp}d)\;, \label{eq:exponents-theta-1} \\
\tilde{\Omega}_{\theta_{\nu P},n} &\equiv& \exp\left[\frac{\bar{\xi}_{\theta_{\nu P},n}}{2}\left(\gamma-\int_{1/L}^{\infty}\frac{2\,dq}{q\left(e^{\hbar v_{{\rm dw}}q/k_B T}-1\right)}\right)\right]\;, \label{eq:exponents-theta-2}\\
\bar{\xi}_{\theta_{\nu P},n}& \equiv& \int^\pi_{-\pi} \frac{d(q_\perp d)}{2\pi }\frac{1}{\tilde{K}_{\nu P}(q_{\perp})}\left[1-\cos(nq_{\perp}d)\right]=\bar{\Delta}_{\theta_{\nu P},n=0}-\bar{\Delta}_{\theta_{\nu P},n}\;  \label{eq:exponents-theta-3} \; .
\end{eqnarray}
\end{subequations}

\section{Kondo-type interaction and RKKY interaction in the domain wall network}
\label{app:RKKY}

\begin{figure}[ht]
    \centering
    \includegraphics[width=0.9\textwidth]{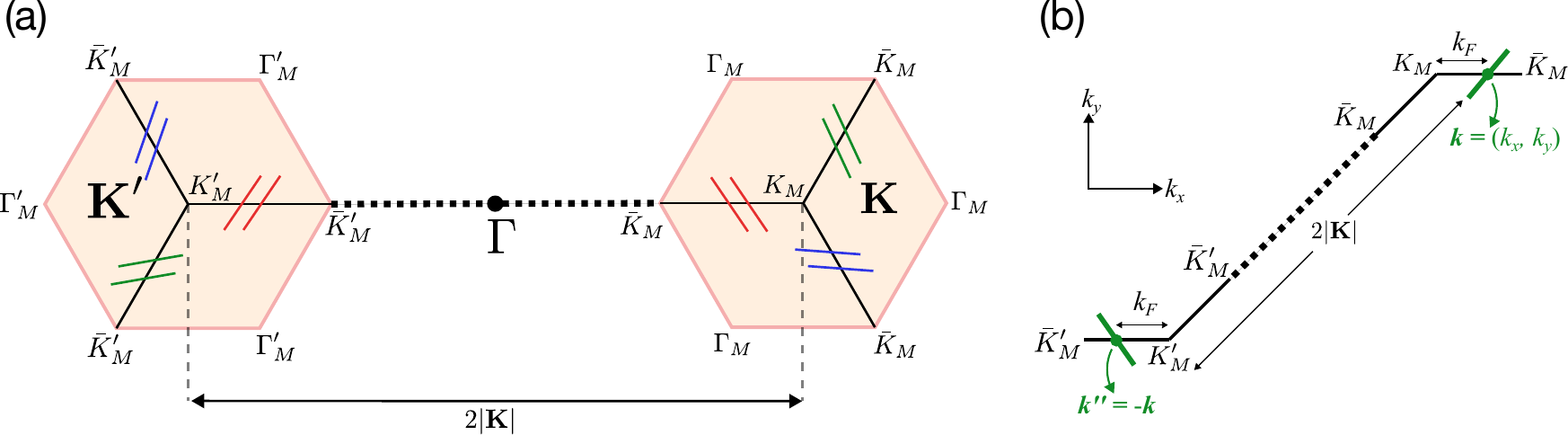}
    \caption{(a) Schematic plot illustrating the \moire Brillouin zones at the $\bm{K}$ (right) and $\bm{K}^\prime$ (left) valleys, each of which hosts domain wall modes (blue, green, red lines) along the three $K_M$-$\bar{K}_M$ and $K_M^\prime$-$\bar{K}_M^\prime$ high-symmetry lines. Here, $\Gamma$ represents   the Brillouin zone center of TBG. (b) Schematic of intervalley backscattering. Here we select the green domain wall modes as an example to illustrate the intervalley backscattering. For convenience, we rotate the coordinates of the momentum space.}
    \label{fig:intervalley}
\end{figure}

In this section, we describe the Kondo-type interaction  in the domain wall network, before discussing the resulting RKKY interaction. In the network, the conduction electron spins couple to the localized moments through 
\begin{align}
   H_{{\rm K}}=& \sum_{k,m}\sum_{\mu , \sigma , \sigma^{\prime}} \frac{J_K^{\mu} }{N_{\perp}} \left[ 
   \psi^{\dagger} _{\sigma,m}(r_{k}) \sigma^\mu_{\sigma\sigma^{\prime}}
   \psi_{\sigma^\prime,m}(r_{k}) \right]   S^\mu_{m} (r_{k}) \;,
   \label{eq:H_K}
\end{align}  
with the effective coupling strength $J_K^{\mu}$ with $\mu \in \{x,y,z\}$, as defined in Eq. (4) in the main text. Distinct from a typical Kondo lattice problem, here we have an electron subsystem described by SLL, which allows us to incorporate the correlation between the electrons as well.
Moreover, the backscattering properties of the electrons traversing the domain walls also affect the transverse components $J_K^{x,y}$ of the effective Kondo-type coupling. 
Namely, the counter-propagating domain-wall modes in Eq.~(\ref{eq:bosonized-fermion}) are constituted by electrons at different valleys, labeled as $\bm{K}$ and $\bm{K}^\prime \equiv -\bm{K}$, in the momentum space of the overall 2D system, as shown in  Fig. \ref{fig:intervalley}(a).  Due to the momentum difference, intervalley backscattering is suppressed as a result of the reduced overlap of the domain-wall electron wave functions. This suppression, in turn, reduces the Kondo-type coupling. 

To quantify the resulting backscattering strength, we calculate the transition amplitude for two counter-propagating domain wall modes on the same domain wall in the presence of a scattering potential through a second-order process. The backscattering process involves two counter-propagating modes on a given domain wall illustrated in Fig. \ref{fig:intervalley}(b), each of which are projected out of the states characterized by the 2D wavevectors  $\bm{k} =(k_{x},k_{y})$ measured from $\Gamma$ and $\bm{k}^{\prime\prime} \equiv -\bm{k}$, with $k_{x},k_{y}>0$, respectively. 
To be precise, we denote the domain wall mode associated with the \(\bm{K}\) valley as \(\left|\bm{k}, m\right\rangle\) and the mode residing in the \(\bm{K}^{\prime}\) valley as \(\left|\bm{k}^{\prime}, m\right\rangle\). Both modes share the same energy $E_{\bm k}$. 

To proceed, we express the second-order transition amplitude under the influence of the potential as 
\begin{align}
    c_{\left|\bm{k},m\right\rangle \to\left|\bm{k}^{\prime\prime},m\right\rangle }^{(2)}&  =\left(\frac{-i}{\hbar}\right)^{2}\int dt_{1}dt_{2}\sum_{\bm{k}^{\prime},m^{\prime}}e^{i\omega_{\bm{k},\bm{k}^{\prime}}t_{1}}V_{\bm{k}m,\bm{k}^{\prime}m^{\prime}}e^{i\omega_{\bm{k}^\prime,\bm{k}}t_2}V_{\bm{k}^{\prime}m^\prime,\bm{k}^{\prime\prime}m}\nn
    & \sim \int d E_{\bm{k}^\prime} \rho(E_{\bm{k}^\prime}) \frac{V_{\bm{k}m,\bm{k}^{\prime}m^{\prime}}V_{\bm{k}^{\prime}m^\prime,\bm{k}^{\prime\prime}m}}{\left(E_{\bm{k}} - E_{\bm{k}^\prime} \right)\left(E_{\bm{k}^\prime}-E_{\bm{k}} \right)}\;,
    \label{eq:c-2}
\end{align}
where $\rho(E_{\bm{k}^\prime})$ denotes the density of states and
\begin{align}
V_{\bm{k}m,\bm{k}^{\prime}m^{\prime}}\equiv\left\langle \bm{k},m\right|\hat{V}\left|\bm{k}^{\prime},m^{\prime}\right\rangle \quad \textrm{and} \quad \omega_{\bm{k},\bm{k}^{\prime}}\equiv\frac{E_{\bm{k}}-E_{\bm{k}^{\prime}}}{\hbar}.
\end{align}
The transition amplitude $ c_{\left|\bm{k},m\right\rangle \to\left|\bm{k}^{\prime\prime},m\right\rangle}^{(2)}$ of Eq.~\eqref{eq:c-2} is inversely dependent on the energy difference of the virtual scattering processes. 
In general, the scattering potential in this higher-order process can arise from the Kondo-type  coupling described in Eq.~\eqref{eq:H_K}, as well as its combination with electron-electron interactions~\cite{Hsu:2020}. To make a conservative estimate, we only consider the smaller contribution from the exchange coupling.
Similarly, we restrict our analysis to the leading virtual processes involving scattering into the intermediate states within the same band as that of  $\left|\bm{k},m\right\rangle$ and $\left|\bm{k}^{\prime\prime},m\right\rangle$, while neglecting the subdominant transitions into the remote bands. 

As a result, the range of the energy integral in Eq.~\eqref{eq:c-2} is constrained by the bandwidth $\Delta_a$. The order of magnitude of the transition amplitude can be estimated by applying the following approximations: $\int d E_{\bm{k}^\prime} \sim \Delta_a$, $\rho(E_{\bm{k}^\prime}) \sim 1/\Delta_a$, $V_{\bm{k}m,\bm{k}^{\prime}m^{\prime}}\sim V_0$ the energy scale of the exchange  coupling, and $|E_{\bm{k}}-E_{\bm{k}^{\prime}}| \lesssim \Delta_a$, yielding the following result: 
\begin{align}
     c_{\left|\bm{k},m\right\rangle \to\left|\bm{k}^{\prime\prime},m\right\rangle }^{(2)} \sim \frac{V_0^2}{\Delta_a^2}\;.
     \label{eq:c-2-final}
\end{align}
This indicates that the backscattering strength is suppressed by an order of $V_0^2/\Delta_a^2$ for domain wall modes, which will also affect the transverse components of Kondo-type coupling when comparing to graphene or carbon nanotube systems. 
Using $\Delta_a \approx 2\sim 10\;\mathrm{meV}$ and Eq.~(\ref{eq:c-2-final}), we estimate the reduced coupling strengths $J_K^{x,y} = O(\mathrm{meV})$ for typical adatoms, based on the experimental value of $5\;\mathrm{meV}$ for fluorine adatoms~\cite{F-adatom-graphene}, and $J_K^{x,y} = O(\mu\mathrm{eV})$ for $^{13}$C, based on the experimental value of $100\; \mu\mathrm{eV}$ in Ref.~\cite{Churchill2009-naturephysics}.
We also remark on the effect of the \moire potential on the effective hyperfine coupling. 
While such investigation is absent, we expect that the misalignment between graphene layers can influence the overlap between $p$-orbital electron wave functions and carbon nuclei at the opposite layers, thereby influencing the Fermi contact and dipolar hyperfine couplings~\cite{Schliemann2003}.

Having described the Kondo-type interaction, we derive the RKKY interaction within the domain wall network by focusing on the weak-$J_{K}$ regime. 
We remark that while the RKKY interaction is a cornerstone in studies of hybrid systems combining conduction electrons with localized spins, its analysis in SLL settings has been insufficient. To fill this gap, we extend the method described in Refs.~\cite{Braunecker-helix-2008,2018-CHH-nuclear-2DTI} to a quasi-2D network in TBG. To this end, we perform the Schrieffer-Wolff transformation on $H = H_{\rm ee} + H_{\rm K}$, and truncate it in the second order.
In momentum space, this procedure leads to   
\begin{align}
    H_{\rm R} = & 
    \frac{1}{N_{{\rm dw}}N}\sum_{q,q_{\perp}}\sum_{\mu,\nu}S_{q,q_{\perp}}^{\nu}\frac{J_{q_{\perp}}^{\mu\nu}(q)}{N_{\perp}^{2}}S_{-q,-q_{\perp}}^{\mu} \;, 
\end{align}
where the RKKY coupling  $J_{q_{\perp}}^{\mu\nu}(q)  \equiv\frac{J_{K}^{2}a^{2}}{2}\chi_{q_{\perp}}^{\mu\nu}(q)$ is determined by the momentum-dependent static spin susceptibility $\chi_{q_{\perp}}^{\mu\nu}(q)$ of the conduction electrons through   
\begin{align}
\chi_{q_{\perp}}^{\mu\nu} (q)&\equiv\frac{-i}{N_{{\rm dw}}N\hbar a^{2}}\lim_{\eta\to0^{+}}\int_{0}^{\infty}dte^{-\eta t}   \left\langle \left[s_{q,q_{\perp}}^{\mu}(t),\,s_{-q,-q_{\perp}}^{\nu} (t=0)\right]\right\rangle_{\rm ee}\;.
\end{align}
The real space representation for $\chi_{q_{\perp}}^{\mu\nu}(q)$ can be readily obtained as
\begin{align}
    \chi_{n}^{\mu\nu}(r)& =\frac{-i}{\hbar}\lim_{\eta\to0^{+}}\int_{0}^{\infty}dte^{-\eta t}  \left\langle \left[s_{n}^{\mu}(r,t),\,s_{n=0}^{\nu}(0,0)\right]\right\rangle _{{\rm ee}} \;, 
    \label{eq:susceptiblity-continuous}
\end{align}
with  $s_{n}^{\mu}(r,t) =  s_{n}^{\mu}(r_{k},t)/a$ at the continuum limit. In terms of domain wall modes, the spin density operators shown in  Eq. (\ref{eq:susceptiblity-continuous}) can be expressed as
\begin{align}
    s_{m}^{\mu}(r) = \frac{1}{2}\sum_{\sigma\sigma^{\prime}}\sum_{\ell \ell^\prime}\sum_{\delta\delta^\prime}\psi_{\ell\delta\sigma,m}^{\dagger}(r)\sigma^\mu_{\sigma\sigma^{\prime}}\psi_{\ell^\prime \delta^\prime \sigma^\prime,m}(r) .
\end{align}
For electronic subsystem respecting the spin rotational symmetry, we have isotropic susceptibility, $\chi_{n}^{\mu\nu} = \delta_{\mu\nu} \chi_{n}^{\mu}$.  
As compared to a single-channel system, the spin susceptibility in the coupled-domain-wall system contains electron correlations across parallel domain walls. These inter-domain-wall correlations manifest as spatial phase coherence between different domain walls, which stabilizes a quasi-2D spin helix ordering.

The integrand in Eq. (\ref{eq:susceptiblity-continuous}) can be computed following the standard procedure~\cite{Giamarchi2003} and expressed in terms of the spin-spin correlator.
The result is a sum of contributions from different momentum transfers due to electronic backscattering,
\begin{align}
    \chi_{n}^{\mu}(r,\tau) =\frac{-1}{4(\pi a)^{2}}
    \left[ \bar{\chi}_{n,2k_{F_{1}}}^{\mu} (r,\tau)
    + \bar{\chi}_{n,2k_{F_{2}}}^{\mu} (r,\tau) +2\bar{\chi}_{n,2Q_{F}}^{\mu} (r,\tau)\right]\;, 
    \label{eq:chi_n-3}
\end{align}
in the imaginary-time form. Note that the factor of 2 in front of $\bar{\chi}^\mu_{n,2Q_F}$ in Eq. (\ref{eq:chi_n-3}) is crucial for determining the helix transition temperature, which will be explained in the following section. At finite temperatures, $\bar{\chi}_{n,2Q}^{\mu}$ takes the form
\begin{align}
    \bar{\chi}_{n,2Q}^{\mu}(r,\tau)&\equiv\cos(2Qr)\tilde{\Omega}_{n,\mathcal{B}}^{\mu}\left(\frac{a}{L}\right)^{\bar{\xi}_{n,\mathcal{B}}^{\mu}/2}  \frac{\left(\frac{\pi k_B T a}{\hbar v_{{\rm dw}}}\right)^{2g_{n,\mathcal{B}}^{\mu}}}{\left[\sinh^{2}\left(\frac{\pi k_B T r}{\hbar v_{{\rm dw}}}\right)+\sin^{2}\left(\frac{\pi k_B T \tau}{\hbar}\right)\right]^{g_{n,\mathcal{B}}^{\mu}}}\;,
    \label{eq:chi-bar-full}
\end{align}
with $\mathcal{B} \in \lbrace \intra,\inter \rbrace$ referred to the $\lbrace\rm{intrabranch},\, \rm{interbranch}\rbrace$ scattering, and 
\begin{align}
    g_{n,\mathcal{B}}^{\mu}&=\begin{cases}
g_{n,{\intra}}^{\mu}, & \text{if }Q=k_{F_{1}}\textrm{ or }k_{F_{2}} , \\
g_{n,{\inter}}^{\mu}, & \text{if }Q=Q_{F}\equiv \left(k_{F_1}+k_{F_2}\right)/2\;,
\end{cases}\; \nn
\tilde{\Omega}_{n,\mathcal{B}}^{\mu}&=\exp\left[\frac{\bar{\xi}_{n,\mathcal{B}}^{\mu}}{2}\left(\gamma-\int_{1/L}^{\infty}\frac{2dq}{q\left(e^{\beta\hbar v_{{\rm dw}}q}-1\right)}\right)\right]
\approx  \exp\left[\frac{\gamma\bar{\xi}_{n,\mathcal{B}}^{\mu}}{2}\right]\; ,\nn
\bar{\xi}_{n,\mathcal{B}}^{\mu}&\equiv4\left(g_{\parallel,\mathcal{B}}^{\mu}-g_{n,\mathcal{B}}^{\mu}\right)\; .
\end{align}
Here, we define $g_{n=0,\mathcal{B}}^{\mu}=g_{\parallel,\mathcal{B}}^{\mu}$, and one can check that  $\tilde{\Omega}_{n=0,\mathcal{B}}^{\mu}=1$ and $ \bar{\xi}_{n=0,\mathcal{B}}^{\mu}=0$ in this case. The exponents $g_{n,\mathcal{B}}^{\mu}$ are  given by
\begin{align}
   g_{n,{\inter}}^{z}&=\frac{1}{4}\left( \bar{\Delta}_{\phi_{cS},n}+\frac{1}{K_{cA}}+K_{sS}+\frac{1}{K_{sA}}\right) \;, \nn
    g_{n,{\inter}}^{x,y}&=\frac{1}{4}\left( \bar{\Delta}_{\phi_{cS},n}+\frac{1}{K_{cA}}+\frac{1}{K_{sS}}+K_{sA}\right) \;,
\end{align}
and 
\begin{align}
   g_{n,{\intra}}^{z}&=\frac{1}{4}\left( \bar{\Delta}_{\phi_{cS},n}+K_{cA}+K_{sS}+K_{sA}\right ) \;, \nn
   g_{n,{\intra}}^{x,y}&=\frac{1}{4}\left( \bar{\Delta}_{\phi_{cS},n}+K_{cA}+\frac{1}{K_{sS}}+\frac{1}{K_{sA}}\right) \; ,
\end{align}
where $\bar{\Delta}_{\phi_{\nu P},n}$ is defined in Eq. (\ref{eq:exponents-1}).

We consider a system with SU(2) symmetry and noninteracting spin sector, leading to $K_{sS}=K_{sA}=1$ and isotropic spin susceptibility ($g_{n,\mathcal{B}}^{x,y}=g_{n,\mathcal{B}}^{z}$). Additionally, we assume a noninteracting charge antisymmetric sector, where $K_{cA} = 1$, resulting in 
\begin{align}
    &g_{n,{\intra}}^{\mu}=g_{n,{\inter}}^{\mu} \equiv g^\mu_n =\frac{1}{4}\left[3+\int_{-\pi}^{\pi}\frac{d(q_{\perp}d)}{2\pi}\cos(nq_{\perp}d)\tilde{K}_{cS}(q_{\perp})\right]\;    ,
    \label{eq:g_n-SU2}
\end{align}
 where $g_{n\geq3}^{\mu}$ in Eq. (\ref{eq:g_n-SU2}) is independent of the inter-domain-wall couplings  $\lambda_{1,2}$ since we only truncate the SLL parameters to the second order. 

After performing Fourier transform on Eq. (\ref{eq:chi-bar-full}), we obtain the momentum-dependent static spin correlator~\cite{Giamarchi2003}, 
    \begin{align}
        \left[\chi_{n}^{R}(q)\right]^{\mu} & =-\tilde{\Omega}_{n}^{\mu}\left(\frac{a}{L}\right)^{\bar{\xi}_{n}^{\mu}/2} \frac{4\sin\left(\pi g_{n}^{\mu}\right)}{(4\pi)^{2}\hbar v_{{\rm dw}}}\left(\frac{\lambda_{T}}{2\pi a}\right)^{2-2g_{n}^{\mu}}\sum_{\eta=\pm}\Bigg\{2\left|B\left(\frac{g^{\mu}_n}{2}-i\frac{\lambda_{T}}{4\pi}\left(q-2\eta Q_{F}\right);1-g_{n}^{\mu}\right)\right|^{2} \nn
        & \qquad\qquad+\sum_{\delta}\left|B\left(\frac{g^{\mu}_{n}}{2}-i\frac{\lambda_{T}}{4\pi}\left(q-2\eta k_{F_{\delta}}\right);1-g_{n}^{\mu}\right)\right|^{2}\Bigg\} \; ,
    \end{align}
where we define $\lambda_T \equiv \hbar v_{\rm dw}/k_B T$ as the thermal length  and 
\begin{align}
B(K;K^\prime)=\frac{\Gamma(K)\Gamma(K^\prime)}{\Gamma(K+K^\prime)}\;.
\end{align}

For later use in calculating the magnon spectrum, we need its Fourier component  $\left[\chi_{n}^{R}(q)\right]^{\mu}$  in the direction perpendicular to the domain walls, $\left[\chi_{q_{\perp}}^{R}(q)\right]^{\mu}=\sum_{n=-(N_{{\rm dw}}-1)}^{N_{{\rm dw}}-1}e^{-iq_{\perp}nd}\left[\chi_{n}^{R}(q)\right]^{\mu}=\sum_{n=0}^{N_{{\rm dw}}-1}\cos\left(q_{\perp}nd\right)\left[\chi_{n}^{R}(q)\right]^{\mu}$, where $\left[\chi_{q_{\perp}}^{R}(q)\right]^{\mu}$ is explicitly given by 
\begin{align}
   \left[\chi_{q_{\perp}}^{R}(q)\right]^{\mu} &  =  \underset{{\rm within~ a~ domain~ wall}}{\underbrace{\left[\chi_{\parallel}^{R}(q)\right]^{\mu}}}+\cos\left(q_{\perp}d\right)\left[\chi_{1}^{R}(q)\right]^{\mu} +\cos\left(2q_{\perp}d\right)\left[\chi_{2}^{R}(q)\right]^{\mu}   +\sum_{n=3}^{N_{{\rm dw}}-1}\cos\left(nq_{\perp}d\right)\underset{\textrm{indep. of }n>3}{\underbrace{\left[\chi_{3}^{R}(q)\right]^{\mu}}}\; .
\end{align}
The coefficient in front of $\left[\chi_{3}^{R}(q)\right]^{\mu}$ can be analytically obtained as
\begin{align}
    \sum_{n=3}^{N_{{\rm dw}}-1}\cos\left(nq_{\perp}d\right) & =\frac{1}{2\sin\left(q_{\perp}d/2\right)}\left\lbrace\sin\left[\left(N_{{\rm dw}}-\frac{1}{2}\right)q_{\perp}d\right]  -\sin\left(\frac{5}{2}q_{\perp}d\right)\right\rbrace \; .
\end{align}
In addition to the intra-domain-wall component $\chi^R_\parallel$, the static spin-spin correlator $\chi^R_{q_\perp}(q)$ also includes contributions from inter-domain-wall spin correlations for $n \neq 0$. These  contributions enhance the RKKY coupling, resulting in a further increase in the helix transition temperature $T_{{\rm hx}}$ compared to the case without the inter-domain-wall correlations, thereby stabilizing the spin helix phase.

\begin{figure}
    \centering
    \includegraphics[width=0.5\textwidth]{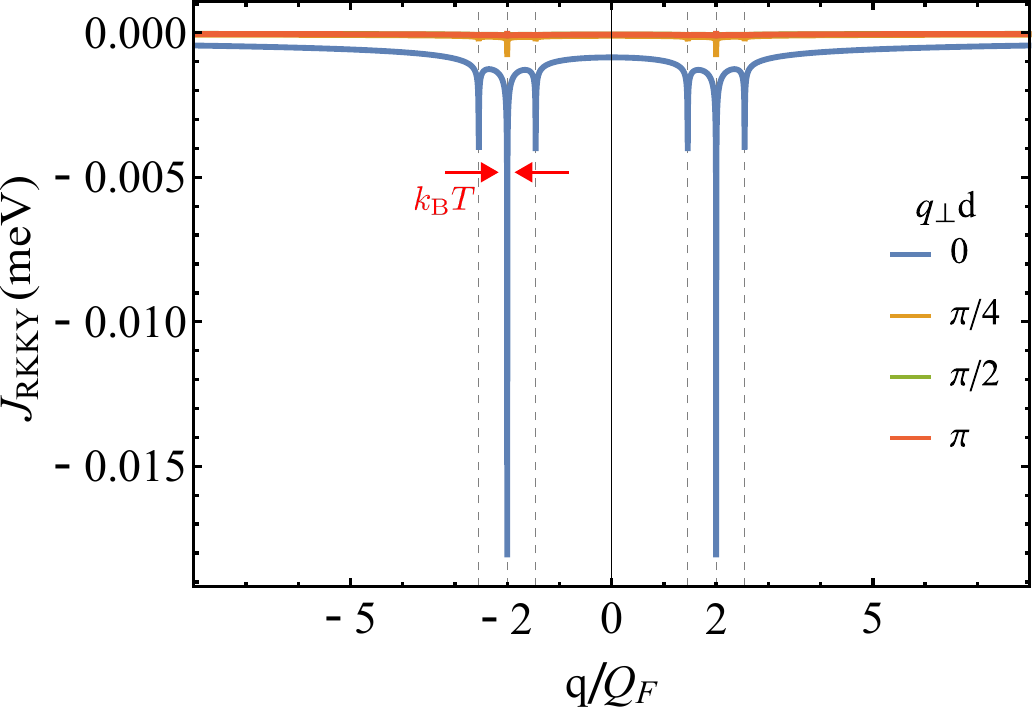}
    \caption{RKKY coupling $J_{q_\perp}(q)$ as a function of $q/Q_F$. Two dips at $q= \pm 2Q_F$ and $q_\perp =0$ can be clearly seen. The other parameters are $T = 0.1$K, $k_{F_1} = 4\times10^8\;\rm{m}^{-1}$,  $k_{F_2} = 7\times10^8\;\rm{m}^{-1}$, $J_K = 1$ meV, $\lambda_1= \sqrt{2}$, $\lambda_2 = 1$, and $N_{\rm{dw}} = 20$. 
    }
    \label{fig:enter-label}
\end{figure}

\section{Formation of quasi-2D spatially phase-coherent spin helix }\label{app:Thx}

In this section, we demonstrate that the RKKY interaction results in formation of a quasi-2D spin helix. To this end, we perform the spin wave analysis at low temperatures.
Motivated by the largest absolute magnitude of the RKKY coupling at the inter-branch scattering momentum, $2Q_{F}$, we consider the following ansatz for the localized spins on a given domain wall, 
\begin{align}
    \langle \bm{S}_{m} (r_{k}) \rangle  & =m_{2Q_{F}}SN_{\perp}\left[\hat{x}\cos\left(2Q_{F}r_{k}+\vartheta_{m}\right)
    +\hat{y}\sin\left(2Q_{F}r_{k}+\vartheta_{m}\right)\right] \;, 
\end{align}
 with its period of oscillation given by $\pi/Q_{F}$. 
 
Without loss of generality,  we choose the helical (spin quantization) axis ($z$ direction) to be parallel to the domain wall, and localized spins are lying on the $xy$ plane, perpendicular to the helical axis. For convenience, we rotate the original spatial coordinate ($\hat{x},\hat{y},\hat{z}$) to a new basis ($\hat{e}_{k,m}^{1},\hat{e}_{k,m}^{2},\hat{e}_{k,m}^{3}$),
\begin{align}
    \begin{pmatrix}
\hat{e}_{k,m}^{1}\\
\hat{e}_{k,m}^{2}\\
\hat{e}_{k,m}^{3}
\end{pmatrix}=\begin{pmatrix}\cos\left(2Q_{F}r_{k}+\vartheta_{m}\right) & \sin\left(2Q_{F}r_{k}+\vartheta_{m}\right) & 0\\
-\sin\left(2Q_{F}r_{k}+\vartheta_{m}\right) & \cos\left(2Q_{F}r_{k}+\vartheta_{m}\right) & 0\\
0 & 0 & 1
\end{pmatrix}
\begin{pmatrix}
\hat{x}\\
\hat{y}\\
\hat{z}
\end{pmatrix}, 
\end{align}
such that $\bm{S}(r_{k})=m_{2Q_{F}}N_{\perp}S\hat{e}_{k}^{1}$. 
In the rotated coordinates, the RKKY interaction is expressed as
\begin{align}
    H_{\rm{R}}	
	=\frac{1}{N_{\perp}^{2}}\sum_{k,l}\sum_{\tilde{\mu}\tilde{\nu}=1,2,3}\sum_{m,n}J_{n}^{\tilde{\mu}\tilde{\nu}}(r_{kl})S_{m+n}^{\tilde{\mu}}(r_{k})S_{m}^{\tilde{\nu}}(r_{l})\;.
 \label{eq:H-RKKY-rotated}
\end{align}
The RKKY couplings in the un-rotated and rotated coordinates, $J_{n}^{\mu}$ and $J_{n}^{\tilde{\mu}\tilde{\nu}}$ are related by
 \begin{align}
     J_{n}^{33}(r_{kl})&=J_{n}^{z}(r_{kl})\;, \nn
     J_{n}^{11}(r_{kl})&=J_{n}^{22}(r_{kl})=J_{n}^{x}(r_{kl})\cos\left(2Q_{F}r_{kl}+\Theta_{n}\right)\;, \nn
     J_{n}^{12}(r_{kl})&=-J_{n}^{21}(r_{kl}) =J_{n}^{x}(r_{kl})\sin\left(2Q_{F}r_{kl}+\Theta_{n}\right)\;,
 \end{align}
where $\Theta_{n}\equiv\vartheta_{m+n}-\vartheta_{m}$ being the phase difference   depending solely on the domain wall separation $\propto n$.

\subsection{Development of phase coherence of spin helices}
\label{app:phase-coherence}
In this subsection, we demonstrate that the RKKY energy at zero temperature is minimized by the configuration of uniform offset phases across parallel domain walls, leading to the formation of a phase-coherent quasi-2D spin helix. 

The RKKY energy resulting from the formation of spin helices in a given array, accounting for inter-domain-wall correlations, can be estimated as
\begin{align}
    E_{{\rm R}}= & \frac{1}{N_{\perp}^{2}}\sum_{m,n}\sum_{k}J_{n}(r_{k})\left\langle \bm{S}_{m+n}(r_{k})\right\rangle \cdot\left\langle \bm{S}_{m}(0)\right\rangle  
    = \frac{a}{N_{\perp}^{2}}\sum_{m,n}\int^L_{-L} dxJ_{n}(r)\left\langle \bm{S}_{m+n}(r)\right\rangle \cdot\left\langle \bm{S}_{m}(0)\right\rangle \;,
\end{align}
with, again, $J_{n}(r)=\left(J_{K}^{2}a/2\right)\chi_{n}(r)$. At zero temperature, the spin susceptibility $\chi_{n}(r)$ is proportional to
\begin{align}
    \chi_{n}(r)\sim - \cos(2Q_{F}r)\left|\frac{a}{r}\right|^{\bar{\Delta}_{n}/2},
\end{align}
with $\bar{\Delta}_{n}$ denoting the exponent depending on LL parameters and $\bar{\Delta}_{n}/2 \geq 1$ for our case. It is straightforward to show that 
    \begin{align}
    E_{{ \rm R }}&\sim 
    \left(-\frac{S^{2} N_{\rm dw}J_{K}^{2}a}{4}\right)\sum_{n}\int^L_{-L} dr\left|\frac{a}{r}\right|^{\bar{\Delta}_{n}/2}\left[\cos\left(4Q_{F}r+\Theta_{n}\right)+\cos\Theta_{n}\right] 
    \nn
   & =  \left(-\frac{S^{2}J_{K}^{2}a}{4}\right) \sum_n C_n \cos \Theta_n ,
    \label{eq:E_R-2}
\end{align}
with $C_n >0$ being a coefficient which we calculate next. The first integral in Eq. (\ref{eq:E_R-2}) leads to
\begin{align}
     \int_{-L}^{L}dr\left|\frac{a}{r} \right|^{\alpha}\cos\left(4Q r+\Theta_{n} \right)  = &  \left(\int^{-a}_{-L} + \int^L_a  \right)dr\left|\frac{a}{r} \right|^{\alpha}\cos\left(4Q r+\Theta_{n} \right) \nn
     = &2 \cos(\Theta_n) \int^L_a dr\left|\frac{a}{r} \right|^{\alpha} \cos(4Qr), 
    \label{eq:C_n-1}
\end{align}
where we impose a small length cutoff $a$ in the integral to avoid   divergence at $r = 0 $. 
Combining the result in Eq. (\ref{eq:C_n-1}) with the other contribution from the second integral in Eq. (\ref{eq:E_R-2}), we reach 
\begin{align}
    C_n = 2 \int^L_a \left|\frac{a}{r} \right|^{\bar{\Delta}_n / 2 } \left[\cos(4Q_F r) + 1 \right] dr > 0\;.
\end{align}
We therefore demonstrate that $C_n > 0$ and that the RKKY energy $E_{{\rm R}}$ is minimized when $\cos\Theta_{n}=1$ (or $\Theta_{n}=0 \;\rm{mod}\;2\pi$). This result suggests phase coherence of the quasi-2D spin helix across parallel domain walls.
Naively speaking, the ``phase-locking''  behavior of the quasi-2D spin helix would facilitate constructive diffraction patterns in neutron scattering measurements~\cite{lovesey-neutron-1984-theory,Helical-spin-theory-NeutronDiff,DyY-ND-PRL,Squires-Neutron-2012,willis-2017-neutron-exp,Furrer-neutron}, distinct from isolated channels such as GaAs~\cite{Braunecker-helix-2008,Braunecker-PRL-2009-C13,Braunecker-2009} or $^{13}$C nanotubes~\cite{CHH-CNT-helix}.
Nevertheless, due to the presence of the metallic gates and 2D nature of the system, such measurements might not be practical. We therefore propose to use resistively-detected spin resonance for the nanoscale systems, as discussed in the main text. 

\subsection{Magnon spectrum}
Now, we proceed to the analysis of magnon spectra. 
Selecting $\hat{e}_{k,m}^{1}$ to be the spin quantization axis, we perform the Holstein–Primakoff transformation on the spin operators  to the magnon operators $a_{km}^{\dagger}(a_{km})$, 
\begin{align}
    S_{m}^{1}(r_{k}) & \approx N_{\perp}S-a_{km}^{\dagger}a_{km}\;, \nn
    S_{m}^{+}(r_{k}) & \approx a_{km}\sqrt{2N_{\perp}S}\;, \nn
     S_{m}^{-}(r_{k}) & \approx a_{km}^{\dagger}\sqrt{2N_{\perp}S}\;.
     \label{eq:magnon-op}
\end{align}
In Eq.~\eqref{eq:magnon-op}, we assume that $N_\perp S$ is large, so that higher-order terms of order $O(1/N_\perp S)$ are subleading and can be neglected. In our case, $N_\perp \approx 80$ and $S = 1/2$, giving $N_\perp S \approx 40$, which is more than an order of magnitude larger than unity and thus justifies this approximation.

\begin{figure*}[ht]
    \centering
    \includegraphics[width=\textwidth]{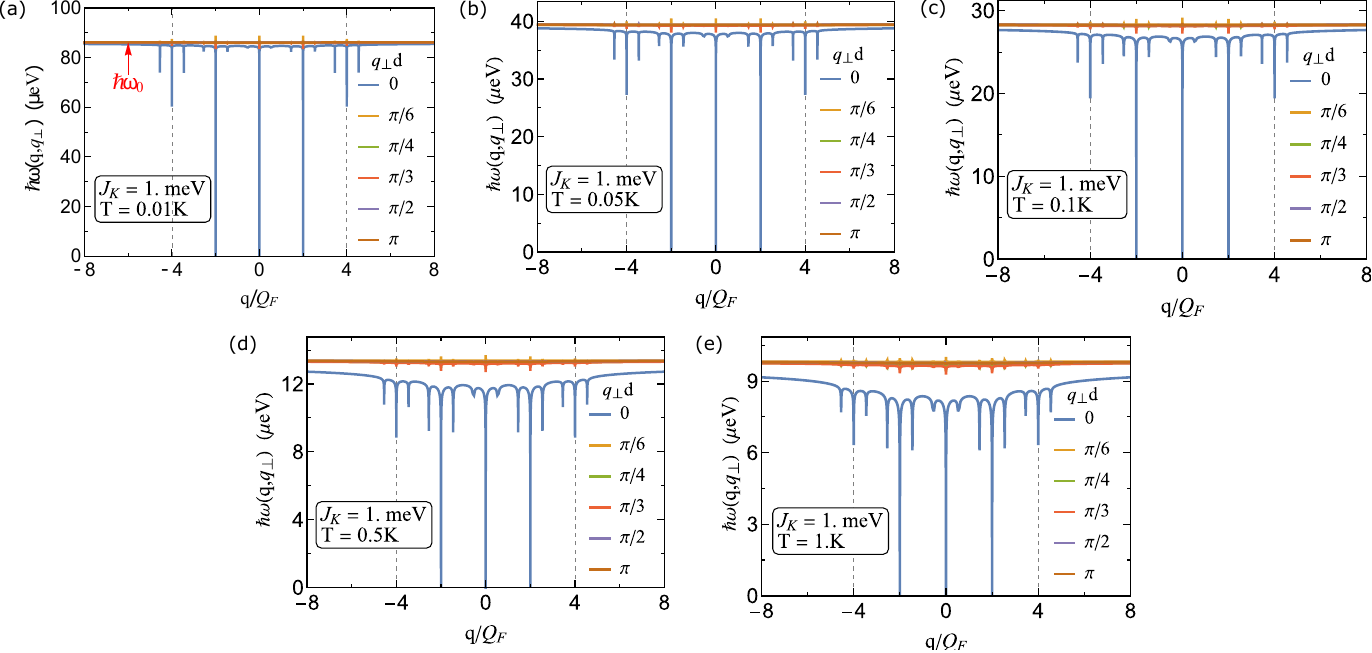}
    \caption{Magnon spectra for different temperatures  $T=0.01,\;0.05,\;0.1,\;0.5,$ and $1$ K, with the exchange coupling fixed at   $J_{K}=1 \;{\rm meV}$ and the number of parallel domain walls set to $N_{{\rm dw}}=20$.  Here, $\hbar \omega_0$ represents the typical scale of the magnon energy.
    }
    \label{fig:magnon-Supp}
\end{figure*}

Since the off-diagonal terms, $H_{\textrm{R}}^{12}$ and $H_{\textrm{R}}^{21}$, in Eq. (\ref{eq:H-RKKY-rotated}) are of odd order in the magnon operators, we have 
$ H_{\textrm{R}}=H_{\textrm{R}}^{11}+H_{\textrm{R}}^{22}+H_{\textrm{R}}^{33}$, 
with 
    \begin{align}
        H_{\textrm{R}}^{11}+H_{\textrm{R}}^{22} = &  \frac{1}{N_{\perp}^{2}}\sum_{k,l}\sum_{m,n}J_{n}^{11}(r_{kl}) \left[\left(N_{\perp}S\right)^{2}-N_{\perp}S\left(a_{k,m+n}^{\dagger}a_{k,m+n}+a_{lm}^{\dagger}a_{lm}\right) \right. \nn
        & \qquad \qquad \left. +\frac{N_{\perp}S}{2}\left(a_{k,m+n}a_{lm}+a_{k,m+n}a_{lm}^{\dagger}+a_{k,m+n}^{\dagger}a_{lm}+a_{k,m+n}^{\dagger}a_{lm}^{\dagger}\right)\right] , \nn
    H_{\textrm{R}}^{33} = & \frac{S}{2N_{\perp}}\sum_{k,l}\sum_{m,n}J_{n}^{z}(r_{kl})\left(-a_{k,m+n}a_{lm}+a_{k,m+n}a_{lm}^{\dagger}+a_{k,m+n}^{\dagger}a_{lm}-a_{k,m+n}^{\dagger}a_{lm}^{\dagger}\right)\;. 
    \end{align}
After the Fourier transform $a_{k,m} = \frac{1}{\sqrt{NN_{\rm{dw}}}}\sum_{q,q_{\perp}}e^{iq_{\perp}y_{m}+iqr_k}a_{q,q_\perp}$, we get  
\begin{align}
    &H_{\textrm{R}}  =\frac{S}{2N_{\perp}}\sum_{q,q_{\perp}} \left[ h_{3}(q,q_{\perp})\left(a_{q,q_{\perp}}^{\dagger}a_{q,q_{\perp}}+a_{-q,-q_{\perp}}a_{-q,-q_{\perp}}^{\dagger}\right)
    + h_{2}(q,q_{\perp})\left(a_{q,q_{\perp}}a_{-q,-q_{\perp}}+a_{-q,-q_{\perp}}^{\dagger}a_{q,q_{\perp}}^{\dagger}\right) \right]\;
\end{align}
with 
\begin{align}
    h_{2}(q,q_{\perp}) & = \frac{1}{2}\left [ J_{q_\perp }^x (2Q_F+q) +J_{q_\perp }^x (2Q_F-q)\right] - J_{q_\perp }^z (q) \;, \nn
h_{3}(q,q_{\perp}) & = -2 J_{q_\perp = 0}^x (2Q_F) + \frac{1}{2}\left [ J_{q_\perp }^x (2Q_F+q) +J_{q_\perp }^x (2Q_F-q)\right] + J_{q_\perp }^z (q)\;.
\end{align} 
Introducing the Nambu basis $\psi^{\dagger}(q,q_{\perp})=\left(a_{q,q_{\perp}}^{\dagger},\,a_{-q,-q_{\perp}}\right)$, we obtain 
\begin{align}
    H_{\textrm{R}}=\frac{S}{2N_{\perp}}\sum_{q,q_{\perp}}\psi^{\dagger}(q,q_{\perp})\mathcal{H}(q,q_{\perp})\psi(q,q_{\perp})  ,
\end{align}
with
\begin{align}
    \mathcal{H}(q,q_{\perp})&=\begin{pmatrix}h_{3}(q,q_{\perp}) & h_{2}(q,q_{\perp})\\
h_{2}(q,q_{\perp}) & h_{3}(q,q_{\perp})
\end{pmatrix}\; .
\end{align}
The eigenvalues of the above bosonic model can be found by $\det\left[\lambda\mathbb{I}-\sigma^{3}\mathcal{H}(q,q_{\perp})\right]=0$, and the resulting spectrum is $\lambda(q,q_{\perp})=\hbar\omega(q,q_{\perp})$ is given in Eq. (7) in the main text. 
We note that the magnon spectrum corresponds to twice the positive eigenvalue of the above matrix, after removing an unphysical, non-positive-definite band. 
In Fig. \ref{fig:magnon-Supp}, we show the magnon spectra for various temperatures and $q_\perp d$ values. Typically, magnon spectra can be detected via resonant inelastic X-ray scattering or inelastic neutron scattering. However, since the metallic gate on top of the TBG sample might make these techniques intractable, we search for alternative features in this work. 

\subsection{Helix ordering temperature $T_{\rm hx}$}
\label{app:feedback}

In this subsection, we estimate the helix transition temperature, denoted as $T_{\rm hx}$, while self-consistently incorporating the static, spatially rotating effective magnetic field generated by the ordered localized spins. Below, we generalize the approach  in Refs.~\cite{Braunecker-2009,CHH-CNT-helix} to two-dimensional network.

When the localized spins are ordered, they induces a static, spatially-rotating Zeeman field that couples back to the conduction-electron spins through $H_{\rm K}$, and changes the conduction-electron spectrum: In the continuum limit, 
$\left\langle \bm{S}_{m}(r)\right\rangle \to\bm{B}_{{\rm hx}}(r)=B_{\rm hx}\left[\hat{x}\cos\left(2Q_{F}r\right)+\hat{y}\sin\left(2Q_{F}r\right)\right]$ with the domain-wall independent field strength $ B_{\rm hx} = m_{2Q_{F}}SJ_{K}$, this induces a  Zeeman term, 
\begin{align}
    H_{{\rm hx}} \equiv \langle H_{\rm K}\rangle_{\rm hx} = \sum_{m}\int dr\bm{B}_{{\rm hx}}(r)\cdot\bm{s}_{m}(r)\;.
    \label{eq:H_hx}
\end{align}
Following the above analysis, we set  $\vartheta_m = 0$. 

To proceed, we split $H_{\rm hx}$ into the interbranch and intrabranch components, $H_{\rm hx} = H_{{\rm hx},\intra} + H_{{\rm hx},\inter}$. In the boson representation, the conduction-electron spin reads
    \begin{align}
        s_{{\inter},m}^{x}(r)&=\frac{1}{2\pi a}\sum_{\delta}\left[\cos\left(-2Q_{F}r+\phi_{cS,m}-\delta\theta_{cA,m}+\delta\phi_{sA,m}-\theta_{sS,m}\right)+\cos\left(-2Q_{F}r+\phi_{cS,m}-\delta\theta_{cA,m}-\delta\phi_{sA,m}+\theta_{sS,m}\right)\right]\;, \nn
        s_{{\inter},m}^{y}(r)&=\frac{1}{2\pi a}\sum_{\delta}\left[\sin\left(-2Q_{F}r+\phi_{cS,m}-\delta\theta_{cA,m}+\delta\phi_{sA,m}-\theta_{sS,m}\right)-\sin\left(-2Q_{F}x+\phi_{cS,m}-\delta\theta_{cA,m}-\delta\phi_{sA,m}+\theta_{sS,m}\right)\right]\;,
    \end{align}
and
    \begin{align}
        s_{{\intra},m}^{x}(r)&=\frac{1}{2\pi a}\sum_{\delta}\left[\cos\left(-2k_{F_{\delta}}r+\phi_{cS,m}+\delta\phi_{cA,m}-\theta_{sS,m}-\delta\theta_{sA,m}\right)+\cos\left(-2k_{F_{\delta}}r+\phi_{cS,m}+\delta\phi_{cA,m}+\theta_{sS,m}+\delta\theta_{sA,m}\right)\right]\;, \nn
        s_{{\intra},m}^{y}(r)&=\frac{1}{2\pi a}\sum_{\delta}\left[\sin\left(-2k_{F_{\delta}}r+\phi_{cS,m}+\delta\phi_{cA,m}-\theta_{sS,m}-\delta\theta_{sA,m}\right)-\sin\left(-2k_{F_{\delta}}r+\phi_{cS,m}+\delta\phi_{cA,m}+\theta_{sS,m}+\delta\theta_{sA,m}\right)\right]\;.
    \end{align}
We have
    \begin{align}
        H_{{\rm hx},{\inter}} \approx\frac{B_{\rm hx}}{2\pi a}\sum_{m,\delta}\int dr\cos\left(\phi_{cS,m}-\delta\theta_{cA,m}-\delta\phi_{sA,m}+\theta_{sS,m}\right)\;,
        \label{eq:H-fb-phi-theta}
    \end{align}
where we have neglected the oscillating term proportional to $\cos\left(-4Q_{F}r+\phi_{cS,m}-\delta\theta_{cA,m}+\delta\phi_{sA,m}-\theta_{sS,m}\right)$ in $ H_{{\rm hx},{\inter}}$ which requires fine-tuning the chemical potential to exactly fit the commensurate condition and is thus not general. The feedback effect $H_{{\rm hx},{\inter}}$ of Eq. (\ref{eq:H-fb-phi-theta}) now shows the same form with the sine-Gordon model, and the modes $\phi_{cS,m}-\delta\theta_{cA,m}-\delta\phi_{sA,m}+\theta_{sS,m}$ will be gapped out when the effective coupling $\sim B_{\rm hx}$ flows to the strong-coupling regime. The RG relevance of this coupling will be addressed below. However, the remaining modes $\phi_{cS,m}-\delta\theta_{cA,m}+\delta\phi_{sA,m}-\theta_{sS,m}$ remains gapless and can still mediate the RKKY coupling.

To better analyze the RKKY interaction mediated by the remaining gapless mode, we construct a new set of boson fields, 
\begin{align}
\Phi_{\delta,m}^{\eta}	= & \frac{1}{2}\left[\phi_{cS,m}-\delta\theta_{cA,m}-\eta\left(\delta\phi_{sA,m}-\theta_{sS,m}\right)\right]\;, \nn
\Theta_{\delta,m}^{\eta}	= & \frac{1}{2}\left[\theta_{cS,m}-\delta\phi_{cA,m}-\eta\left(\delta\theta_{sA,m}-\phi_{sS,m}\right)\right]\;, 
\label{eq:Phi-Theta}
\end{align}
with $\eta \in \lbrace +1, -1 \rbrace$. The  bosons, $\Phi_{\delta}^{\eta}$ and $\Theta_{\delta}^{\eta}$,   obey the standard commutator, $\left[\Phi_{\delta,m}^{\eta}(r),\,\Theta_{\delta^{\prime},m^\prime}^{\eta^{\prime}}(r^\prime)\right] = \frac{i\pi}{2}\text{sgn}(r^\prime-r )\delta_{mm^\prime}\delta_{\delta\delta^{\prime}}\delta_{\eta\eta^{\prime}}$. In the new basis, the electron subsystem of Eq.~(\ref{eq:H_ee-3}) and $H_{{\rm hx},\inter}$ takes the form  
\begin{align}
    H_{{\rm ee}} & = \frac{1}{N_{{\rm dw}}}\sum_{q_{\perp},\delta,\eta}\int\frac{\hbar dr}{2\pi}\left[ \frac{\tilde{v}(q_{\perp})}{\tilde{K}(q_{\perp})}\left|\partial_{r}\Phi_{\delta,q_{\perp}}^{\eta}(r)\right|^{2}+\tilde{v}(q_{\perp})\tilde{K}(q_{\perp})\left|\partial_{r}\Theta_{\delta,q_{\perp}}^{\eta}(r)\right|^{2}\right] \;, \nn
    H_{{\rm hx},{\inter}} & =  \frac{B_{\rm hx}}{2\pi a}\sum_{m,\delta}\int dr\cos\left[2\Phi_{\delta,m}^{+}(r)\right] ,
    \label{eq:Hee-modified}
\end{align}
with a modified velocity and SLL parameters,
\begin{align}
    \tilde{v}(q_{\perp}) \approx \tilde{v}_{\rm dw} =   \frac{v_{{\rm dw}}}{4}\sqrt{3K_{cS}+\frac{3}{K_{cS}}+10} \;, \quad 
   \tilde{K}(q_{\perp}) =\sqrt{\frac{\tilde{K}_{cS}^{2}(q_{\perp})+3\tilde{K}_{cS}(q_{\perp})}{3\tilde{K}_{cS}(q_{\perp})+1}} \, .
   \label{eq:tilde-v-K}
\end{align}
Note that we do not include the marginal terms in the above, and assume $v_{\nu P}(q_{\perp})\approx v_{{\rm dw}}$ in Eq. (\ref{eq:Hee-modified}). In the derivation for $\tilde{v}$ in Eq. (\ref{eq:tilde-v-K}), we assume $\tilde{K}_{cS}(q_{\perp}) \approx K_{cS}$.

\begin{figure*}
    \centering
    \includegraphics[width=0.9\textwidth]{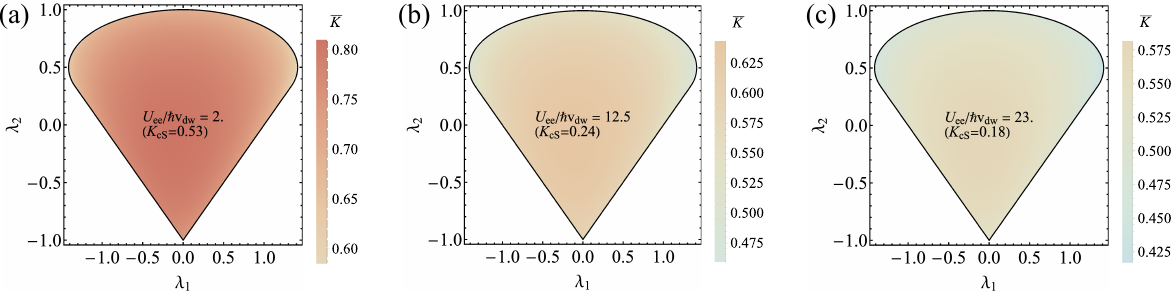}
    \caption{(a)-(c) Effective SLL parameter $\bar{K}$ as a function of $\lambda_1$ and $\lambda_2$  for various values of  electrostatic energy $U_{\rm ee}/\hbar v_{\rm dw} = 2, \; 12.5$ and $23$.}
    \label{fig:Kbar}
\end{figure*}
By computing the zero-temperature correlation function of $\left\langle \cos\left[2\Phi_{\delta,m}^{+}(r)\right]\cos\left[2\Phi_{\delta,m}^{+}(0)\right]\right\rangle $, the scaling dimension of the effective magnetic field $B_{\rm hx}$ is found to be $1-\bar{K}$. For the parameters considered in our investigation, we numerically verify that $1-\bar{K} > 0$ [see Fig. \ref{fig:Kbar}], indicating that the sine-Gordon term $ H_{{\rm hx},{\inter}}$ is always relevant. 

The  contribution to  the susceptibility now only comes from the $\Phi^-_\delta$ field,   whose correlation function across $n$ different domain walls is now characterized by the feedback-modified bandwidth $\tilde{\Delta}_{a}\equiv\hbar \tilde{v}_{{\rm dw}}/a$ and the parameter 
\begin{align}
    \bar{K}_n= \int^\pi_{\pi/N_{\rm dw}} \frac{d(q_\perp d)}{\pi}\cos(nq_\perp d) \tilde{K}(q_\perp)\;.
    \label{eq:Kbar-n}
\end{align}
Here, to regularize the singularity in $\tilde{J}^x_n (q \pm 2Q_F)$ in Eq. (\ref{eq:tilde-J_x}), which arises when $\lambda_1 = \lambda_2 = 0$ (or equivalently $\bar{K}_n = 0$), we introduce a lower cutoff $q_\perp=\pi / (N_{\rm dw}d)$ in the integral over $\bar{K}_n$ in Eq. (\ref{eq:Kbar-n}). Retaining only the $\Phi^-_\delta$ field, we get 
\begin{align}
    \tilde{J}^x_{n}(q = \pm2Q_{F})\approx\frac{-4J_{K}^{2}\sin\left(\pi \bar{K}_{n}\right)}{(4\pi)^{2}\tilde{\Delta}_{a}}\left(\frac{\tilde{\Delta}_{a}}{2\pi k_{B}T}\right)^{2-2\bar{K}_{n}}\left|B\left(\frac{\bar{K}_{n}}{2};1-\bar{K}_{n}\right)\right|^{2},
    \label{eq:tilde-J_x}
\end{align}
which allows us to find $T_{\rm hx}$ below.

\begin{figure}[t]
    \centering
    \includegraphics[width=0.95\textwidth]{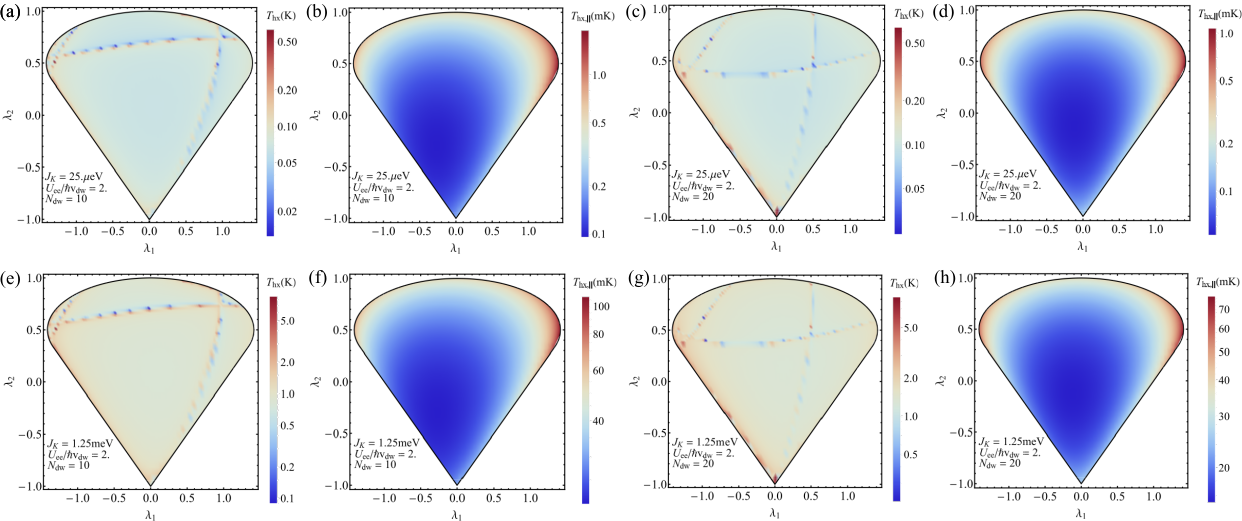}
    \caption{(a,c,e,g) Ordering temperature $T_{\rm{hx}}$  and (c,d,f,h) artificially defined scale $T_{{\rm hx},\parallel}$ as a function of the interaction parameters $\lambda_1$ and $\lambda_2$ for (a)-(d) $J_K = 25$ $\mu\rm{eV}$ and (e)-(h) $J_K = 1.25\;\rm{meV}$, and for (a,b,e,f) $N_{\rm dw} = 10$ and (c,d,g,h) $N_{\rm dw} = 20$.  The  $J_K$ values  are motivated by the realizations by nuclear spins  and magnetic adatoms. The temperature scale $T_{{\rm hx},\parallel}$ accounts only for intra-domain-wall RKKY interaction and serves as a reference for comparison with $T_{\rm hx}$. Here we adopt $U_{\rm ee}/\hbar v_{\rm dw}  =2$. }
    \label{fig:Thx}
\end{figure}

\begin{figure}[ht]
    \centering
    \includegraphics[width=0.95\textwidth]{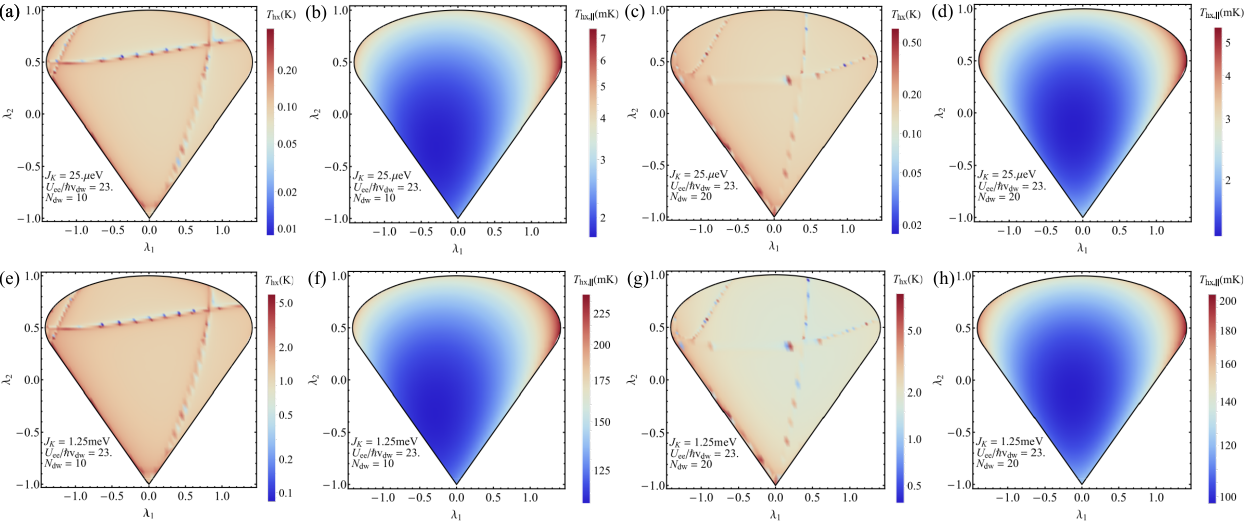}
    \caption{Similar plots to Fig. \ref{fig:Thx}, but for $U_{\rm{ ee}}/\hbar v_{\rm dw}  = 23$. }
    \label{fig:Thx-2}
\end{figure}

We define the ordering temperature as the one where the number of magnons is comparable to the number of spin-flips that randomize the magnetic order, leading to 
\begin{align}
      1- \frac{1/S}{e^{\hbar \omega_0/k_B T_{\rm hx} }-1} =0\;.
    \label{eq:m2QF}
\end{align}
In the above, we have used the fact that in realistic systems, the zero-energy magnon modes are gapped due to finite-size effects~\cite{CHH-CNT-helix,Braunecker-2009}. Specifically, the magnon energy is given by the RKKY energy scale, 
\begin{align}
\hbar\omega_{0}(T_{\rm hx})=\frac{S\left|\tilde{J}^x_{q_\perp = 0} (2Q_{F}, T_{\rm hx})\right|}{2N_{\perp}} \;.
\label{eq:transcedental-eq}
\end{align}
The transcendental equation in Eq. (\ref{eq:transcedental-eq}) can then be numerically solved, with examples shown in panels (a), (c), (e), (g) of Figs. \ref{fig:Thx}  and \ref{fig:Thx-2}. In the allowed values of $\lambda_1$ and $\lambda_2$, we show that the contributions from the inter-domain-wall RKKY coupling enhances the ordering temperature $T_{\rm hx}$ by more than one order of magnitude  as compared with the rather ``unphysical'' temperature scale $T_{{\rm hx},\parallel}$ that excludes such inter-domain-wall contributions. We also note additional features, where the distribution of  $T_{\rm hx}$ shows alternating enhancement and suppression, as demonstrated in panels (a), (c), (e), and (g) of Figs.~\ref{fig:Thx} and \ref{fig:Thx-2}. This non-uniform distribution arises from the summation of varying non-uniformities in $\tilde{J}_{n}^x$ of Eq.~\eqref{eq:tilde-J_x} across different $n$. The more important information is the overall scale of the ordering temperature, which is in the experimentally accessible regime. 
Remarkably, the non-uniformity becomes less pronounced as the number of parallel domain walls $N_{\rm dw}$ increases. More significantly, the overall order of magnitudes of $T_{\rm hx}$ increases as the electron interaction $U_{\rm ee}$ increases, as expected.
 

\section{Electron-magnon interaction and its influence on correlation functions}
 
In this section, we discuss the electron-magnon interaction $H_{{\rm em}}$, which can be introduced through  
\begin{align}
    H_{{\rm K}}=H_{\rm hx}+H_{{\rm em}} \;,
\end{align}
 where the first term $H_{\rm hx}$ has been defined in Eq. (\ref{eq:H_hx}), and the second term is derived through Holstein–Primakoff transformation, giving 
 \begin{align}
     H_{{\rm em}}	& =  \frac{J_K}{N_{\perp}}\sum_{m,k}\Bigg\{-s_{m}^{1}(r_{k})a_{km}^{\dagger}a_{km}   +  \sqrt{\frac{N_{\perp}S}{2}}s_{m}^{2}(r_{k})\left(a_{km}+a_{km}^{\dagger}\right)
     + \frac{1}{i}\sqrt{\frac{N_{\perp}S}{2}}s_{m}^{3}(r_{k})\left(a_{km}-a_{km}^{\dagger}\right)\Bigg\rbrace \;.
 \end{align}
 Here, we only focus on the contribution of the small-momentum transfer $q \sim 0$ of $s_{m}^{3} (r_k)$. Upon bosonization, we have 
 \begin{align}
     s_{m}^{3}(r)\approx\frac{-1}{\pi}\partial_{r}\phi_{sS,m}(r)\;, 
     \label{eq:s_z-phi_sS}
 \end{align}
which leads to
 \begin{align}
     H_{{\rm em}}	\approx &\frac{iJ_K}{\pi N_{\perp}}\sqrt{\frac{N_{\perp}S}{2}}\sum_{m,k}\left[\partial_{r_{k}}\phi_{sS,m}(r_{k})\right]\left(a_{km}-a_{km}^{\dagger}\right) 
     =
     g_{{\rm em}}\int\frac{dr}{2\pi}\sum_{m}\left[\partial_{r}\phi_{sS,m}(r)\right]\Pi_{m}(r)\;, 
 \label{eq:H_em}
 \end{align}
 where  we have replaced $\sum_{k}\to  \int dr/a$,  $(a_{km}-a_{km}^{\dagger} ) / \sqrt{a} \to [  a_{m}(r)-a_{m}^{\dagger}(r)]$, neglected the contributions from $S_{m}^{1,2}(r_{k})$, and defined the effective coupling strength  $g_{{\rm em}}\equiv-2J_{K}\sqrt{a S m_{2Q_F}/(\hbar\omega_{0}N_{\perp})}$. 
 In the above, the field  $a_{k,m}$ is dimensionless while $a_{m}(r)$ carries dimension of $a^{-1/2}$. The prefactor $1/2\pi$ has been introduced for convenience. We also introduce the conjugate field $\Pi_{m}(r)$ to  $\varphi_{m}(r)$, 
 \begin{align}
     \Pi_{m}(r)	& = \frac{1}{\sqrt{LN_{{\rm dw}}}}\sum_{q,q_{\perp}}\sqrt{\frac{\hbar}{2\omega_{0}}}\left(e^{iqr+iq_{\perp}y_{m}}a_{q,q_{\perp}}  +e^{-iqr-iq_{\perp}y_{m}}a_{q,q_{\perp}}^{\dagger}\right) \;, \\
     \varphi_{m}(r)	& =\frac{1}{\sqrt{LN_{{\rm dw}}}}\sum_{q,q_{\perp}}\sqrt{\frac{\hbar\omega_{0}}{2}}\left(\frac{1}{i}\right)\left(e^{iqr+iq_{\perp}y_{m}}a_{q,q_{\perp}}   -e^{-iqr-iq_{\perp}y_{m}}a_{q,q_{\perp}}^{\dagger}\right)\;, 
     \label{eq:magnon-phi-Pi}
 \end{align}
with $a_{q,q_{\perp}}$ being dimensionless operator in the momentum space. It is straightforward to verify the fact that $\varphi_{m}$ and $\Pi_{m}$ forms a set of conjugate operators, namely
 \begin{align}
     \left[\varphi_{m}(r),\,\Pi_{m^{\prime}}(r^{\prime})\right]=i\hbar\delta(r-r^{\prime})\delta_{mm^{\prime}}\;.
 \end{align}

As can be seen in Eq.~(\ref{eq:H_em}),  $H_{{\rm em}}$ represents the coupling between the density-like operator $\partial_{r}\phi_{sS,m}$ in the spin-symmetric sector and the conjugate field $\Pi_{m}$. While its derivation is reminiscent of electron-phonon coupling in the charge sector of  different systems~\cite{Bardeen-RMP-WB,Wentzel-PR-WB,Loss-WB,Martin-WB-PRB,CHH-nano-horizon-2024, HCWang-TBG}, there is an important difference. Namely, the magnons here are approximately dispersionless with a constant energy $\hbar \omega_{0}$, as discussed in the main text. This turns out to give a different behavior in the modified scaling dimension in the bosonic operators. 

To proceed, we evaluate the propagators of the $\phi_{\nu P}$ and $\theta_{\nu P}$ fields and the corresponding scaling dimensions. 
Since Eq.~(\ref{eq:H_em}) only involves  $\phi_{sS,m}$ fields, we focus on this sector,   $S_{sS}+S_{{\rm mag}}+S_{{\rm em}}$, with the following terms,
    \begin{align}
        \frac{S_{sS}}{\hbar}&=\frac{1}{N_{{\rm dw}}}\sum_{q_{\perp}}\int\frac{d^{2}r}{2\pi}\Bigg\{-2i\left[\partial_{r}\theta_{sS,q_{\perp}}(\bm{r})\right]\left[\partial_{\tau}\phi_{sS,-q_{\perp}}(\bm{r})\right]
        +\frac{v_{{\rm dw}}}{K_{sS}}\left|\partial_{r}\phi_{sS,q_{\perp}}(\bm{r})\right|^{2}+v_{{\rm dw}}K_{sS}\left|\partial_{r}\theta_{sS,q_{\perp}}(\bm{r})\right|^{2}\Bigg\}\;, \nn
        \frac{S_{{\rm mag}}}{\hbar}&=\frac{1}{N_{{\rm dw}}}\sum_{q_{\perp}}\int\frac{d^{2}r}{2\pi}\Bigg\{-\frac{2\pi i}{\hbar}\Pi_{q_{\perp}}(\bm{r})\left[\partial_{\tau}\varphi_{-q_{\perp}}(\bm{r})\right]+\frac{\pi}{\hbar}\left|\Pi_{q_{\perp}}(\bm{r})\right|^{2}+\frac{\pi\omega_{0}^{2}}{\hbar}\left|\varphi_{q_{\perp}}(\bm{r})\right|^{2}\Bigg\}\;, \nn
         \frac{S_{{\rm em}}}{\hbar}&=\frac{1}{N_{{\rm dw}}}\sum_{q_{\perp}}\int\frac{d^{2}r}{2\pi}\,\frac{g_{{\rm em}}}{\hbar}\left[\partial_{r}\phi_{sS,q_{\perp}}(\bm{r})\right]\Pi_{-q_{\perp}}(\bm{r})\; ,
         \label{eq:S-emeg-relevant}
    \end{align}
where $\bm{r}\equiv(r,\tau)$ and $d^{2}r\equiv drd\tau$. Since Eq. (\ref{eq:S-emeg-relevant}) are quadratic in all the fields, we can derive the effective action for the $\phi_{sS}$ and $\theta_{sS}$ fields nonperturbatively, giving rise to 
    \begin{align}
        \frac{S[\phi_{sS}]}{\hbar}&=\frac{k_B T}{\hbar LN_{{\rm dw}}}\sum_{q_{\perp},\bm{p}}\frac{1}{2\pi K_{sS}}\left[v_{{\rm dw}}q^{2}+\frac{\omega_{n}^{2}}{v_{{\rm dw}}}-\frac{g_{{\rm em}}^{2}\omega_{0}^{2}K_{sS}q^{2}}{4\pi\hbar\left(\omega_{0}^{2}+\omega_{n}^{2}\right)}\right]\left|\phi_{sS,q_{\perp}}(\bm{p})\right|^{2}\;, \\
        \frac{S[\theta_{sS}]}{\hbar}&=\frac{k_B T}{\hbar  LN_{{\rm dw}}}\sum_{q_{\perp},\bm{p}}\frac{K_{sS}}{2\pi}\left[v_{{\rm dw}}q^{2}+\omega_{n}^{2}\frac{4\pi\hbar\left(\omega_{0}^{2}+\omega_{n}^{2}\right)}{4\pi\hbar v_{{\rm dw}}\left(\omega_{0}^{2}+\omega_{n}^{2}\right)-g_{{\rm em}}^{2}\omega_{0}^{2}K_{sS}}\right]\left|\theta_{sS,q_{\perp}}(\bm{p})\right|^{2} \;,
        \label{eq:modified-boson-action}
    \end{align}
with $\bm{p} \equiv (q,\omega_n)$. One can find the two-point correlation functions for the $\phi_{sS}$ and $\theta_{sS}$ fields as 
    \begin{align}
    \left\langle \bar{\phi}_{sS,q_{\perp}}(\bm{p})\phi_{sS,q_{\perp}}(\bm{p})\right\rangle _{{\rm ee}+{\rm em}} & =\frac{\pi\hbar LN_{{\rm dw}}K_{sS}/k_B T}{v_{{\rm dw}}q^{2}+\frac{\omega_{n}^{2}}{v_{{\rm dw}}}-\frac{g_{{\rm em}}^{2}\omega_{0}^{2}K_{sS}q^{2}}{4\pi\hbar\left(\omega_{0}^{2}+\omega_{n}^{2}\right)}}\;, \\
     \left\langle \bar{\theta}_{sS,q_{\perp}}(\bm{p})\theta_{sS,q_{\perp}}(\bm{p})\right\rangle _{{\rm ee}+{\rm em}} & =\frac{\pi\hbar LN_{{\rm dw}}/(k_B T K_{sS})}{v_{{\rm dw}}q^{2}+\omega_{n}^{2}\frac{4\pi\hbar\left(\omega_{0}^{2}+\omega_{n}^{2}\right)}{4\pi\hbar v_{{\rm dw}}\left(\omega_{0}^{2}+\omega_{n}^{2}\right)-g_{{\rm em}}^{2}\omega_{0}^{2}K_{sS}}}\;,
     \label{eq:prop-e-mag}
    \end{align}
where $\langle \cdots \rangle_{\rm ee + em}$ denotes the expectation value with respect to full electron-magnon-coupled system. 

The effective action is characterized by excitation modes whose corresponding energy $ \hbar \omega_{\pm} $ can be extracted from the poles of the propagators in Eq.~\eqref{eq:modified-boson-action}, which can be expressed as  
    \begin{align}
        \ensuremath{\omega_{\pm}^{2}(q)} = \frac{v_{{\rm dw}}^{2}q^{2}+\omega_{0}^{2}\pm\sqrt{\left(v_{{\rm dw}}^{2}q^{2}-\omega_{0}^{2}+2\Gamma^{2}_{\rm em}/v_{{\rm dw}}^{2}\right)^{2}+4\Gamma^{4}_{\rm em}/v_{{\rm dw}}^{4}-4\omega_{0}^{2}\Gamma^{2}_{\rm em}/v_{{\rm dw}}^{2}}}{2}\;,
        \label{eq:disp-e-mag}
    \end{align}
with  
\begin{align}
    \Gamma_{\rm em}\equiv\frac{g_{{\rm em}}\omega_{0}}{2}\sqrt{\frac{v_{{\rm dw}}K_{sS}}{\pi\hbar}}\;.
\end{align}
One can check that for $\Gamma_{\rm em}=0$, $\hbar \omega_{\pm}$  reduces to the original energy for the decoupled $\phi_{sS} \, (\theta_{sS})$ and magnons, $(\hbar \omega_{\pm})^2 \to (\hbar v_{{\rm dw}} q)^2,\; (\hbar \omega_{0})^2$,  respectively. 
We explore the general properties of the excitation energy. 
At $q=0$, one finds that $\omega_{-}$ is gapless while $\omega_{+}$ is gapped, with $\omega_{+} (q=0)=\omega_{0} $. The leading order behavior of $\omega_{\pm}$ in small $q$ can be found as
\begin{align}
    \omega_{+}^{2}(q)\approx\omega_{0}^{2}\;,\quad\omega_{-}^{2}(q)\approx\left(v_{{\rm dw}}^{\prime}\right)^{2}q^{2} , 
    \label{eq:disp-approximate}
\end{align}
with   the modified velocity $v_{{\rm dw}}^{\prime}$ of $\phi_{sS}$ by magnons,  
\begin{align}
    v_{{\rm dw}}^{\prime}\equiv v_{{\rm dw}}\sqrt{1-\Gamma^{2}_{\rm em}/\omega_{0}^{2}v_{{\rm dw}}^{2}}\;.
    \label{eq:vdw-prime}
\end{align}
 
Next, we calculate various correlation functions for $\phi_{sS}$ and $\theta_{sS}$ in the low-energy (small-momentum) regime. In this regime, using Eqs.~(\ref{eq:disp-approximate}) and (\ref{eq:vdw-prime}), we approximate the two-point correlators as
\begin{align}
    &\left\langle \bar{\phi}_{sS,q_{\perp}}(\bm{p})\phi_{sS,q_{\perp}}(\bm{p})\right\rangle _{{\rm ee}+{\rm em}}  \approx \frac{\hbar LN_{{\rm dw}}\pi K_{sS}v_{{\rm dw}}/(k_B T)}{\omega_{n}^{2}+\left(v_{{\rm dw}}^{\prime}\right)^{2}q^{2}} \;, \nn
    & \left\langle \bar{\theta}_{sS,q_{\perp}}(\bm{p})\theta_{sS,q_{\perp}}(\bm{p})\right\rangle _{{\rm ee}+{\rm em}}  \approx \frac{\hbar LN_{{\rm dw}}\pi v_{{\rm dw}}}{k_B T K_{sS}} 
    \left(\frac{1-\Gamma_{{\rm em}}^{2}/v_{{\rm dw}}^{2}\omega_{0}^{2}}{\omega_{n}^{2}+\left(v_{{\rm dw}}^{\prime}\right)^{2}q^{2}}+\frac{\Gamma_{{\rm em}}^{2}/v_{{\rm dw}}^{2}\omega_{0}^{2}}{\omega_{n}^{2}+\omega_{0}^{2}}\right) .
\end{align}
We first evaluate the correlation function for $\phi_{sS}$ of the following form, 
    \begin{align}
    G_{\phi_{sS},m,n}(\bm{r}) & \equiv\left\langle \left[\phi_{sS,m+n}(\bm{r})-\phi_{sS,m}(0)\right]^{2}\right\rangle _{{\rm ee}+{\rm em}} \nn
         &=\int_{-\pi}^{\pi}\frac{d(q_{\perp}d)}{2\pi}\int_{-\Lambda}^{\Lambda}\frac{dq}{2\pi}\,  \frac{2\pi\hbar K_{sS}v_{{\rm dw}}}{2\hbar v_{{\rm dw}}^{\prime}\left|q\right|}\,\Bigg\{ 2n_{B}\left(\hbar v_{{\rm dw}}^{\prime}\left|q\right|\right)+1  \nn
         & \qquad\qquad -\cos\left(qr+q_{\perp}nd\right)\left[2n_{B}\left(\hbar v_{{\rm dw}}^{\prime}\left|q\right|\right)\cosh\left(v_{{\rm dw}}^{\prime}\left|q\right|\tau\right)+e^{-v_{{\rm dw}}^{\prime}\left|q\right|\tau}\right]
\Bigg\} .
\label{eq:em-G-phisS}
    \end{align}
In the above, we introduce the momentum cutoff $\Lambda$, such that $v_{{\rm dw}}^{\prime}\Lambda=\omega_{0}$. This corresponds to a small distance cutoff  $\tilde{a}=1/\Lambda=v_{{\rm dw}}^{\prime}/\omega_{0}$. In the $T=0$ limit, $G_{\phi_{sS},m,n}(\bm{r})$ in Eq. (\ref{eq:em-G-phisS}) becomes
    \begin{align}
        G_{\phi_{sS},m,n}(\bm{r})&=\int_{-\pi}^{\pi}\frac{d(q_{\perp}d)}{2\pi}\int_{0}^{\Lambda}dq\,\frac{K_{sS}v_{{\rm dw}}}{v_{{\rm dw}}^{\prime}}\,\frac{1}{q}\left[1-\cos(qr)\cos(q_{\perp}nd)e^{-v_{{\rm dw}}^{\prime}\left|q\right|\tau}\right] \nn
        & =\begin{cases}
\frac{K_{sS}v_{{\rm dw}}}{v_{{\rm dw}}^{\prime}}\ln\left|\frac{\tilde{r}}{\tilde{a}}\right|\,, & n=0\,,\\
\frac{K_{sS}v_{{\rm dw}}}{v_{{\rm dw}}^{\prime}}\left(-\gamma-\ln\frac{\tilde{a}}{L}\right) & n\neq0\,,
\end{cases}
    \end{align}
with $\tilde{r}=\sqrt{\left(v_{{\rm dw}}^{\prime}\tau+\tilde{a}\right)^{2}+r^{2}}$. 

The correlation function of $\theta_{sS}$ takes the following form, 
    \begin{align}
    G_{\theta_{sS},m,n}(\bm{r}) & \equiv\left\langle \left[\theta_{sS,m+n}(\bm{r})-\theta_{sS,m}(0)\right]^{2}\right\rangle_{\rm{ee+em}} \nn 
    &   =\frac{k_B T}{\hbar LN_{{\rm dw}}}\sum_{\bm{p},q_{\perp}}\frac{\pi v_{{\rm dw}}}{K_{sS}}\left[\frac{1-\Gamma_{{\rm em}}^{2}/v_{{\rm dw}}^{2}\omega_{0}^{2}}{\omega_{n}^{2}+\left(v_{{\rm dw}}^{\prime}\right)^{2}q^{2}}+ \frac{\Gamma_{{\rm em}}^{2}/v_{{\rm dw}}^{2}\omega_{0}^{2}}{\omega_{n}^{2}+\omega_{0}^{2}}\right] \Big[2-2\cos\left(qx+q_{\perp}nd-\omega_{n}\tau\right)\Big] \nn
    &  \equiv I_1 + I_2\;,
    \label{eq:G_theta_sS-em}
    \end{align}
The $I_1$ term in Eq. (\ref{eq:G_theta_sS-em}) above gives 
    \begin{align}
        I_1&= \frac{k_B T}{ LN_{{\rm dw}}}\sum_{\bm{p},q_{\perp}}\frac{\pi v_{{\rm dw}}\left(1-\Gamma_{{\rm em}}^{2}/v_{{\rm dw}}^{2}\omega_{0}^{2}\right)}{K_{sS}}\frac{2\hbar}{-(i\hbar\omega_{n})^{2}+\hbar^{2}\left(v_{{\rm dw}}^{\prime}\right)^{2}q^{2}}\left[1-\cos\left(qr+q_{\perp}nd-\omega_{n}\tau\right)\right] \nn
        & =\begin{cases}
\frac{v_{{\rm dw}}^{\prime}}{K_{sS}v_{{\rm dw}}}\ln\left|\frac{\tilde{r}}{\tilde{a}}\right|\,, & n=0\,,\\
\frac{v_{{\rm dw}}^{\prime}}{K_{sS}v_{{\rm dw}}}\left(-\gamma-\ln\frac{\tilde{a}}{L}\right)\,, & n\neq0\,.
\end{cases}
    \end{align}
We observe that  the correlation within a domain wall exhibits a logarithmic divergence with distance and time, while the correlation across different domain walls remains constant. This difference arises because the spin-symmetric sector is characterized by identical  velocity and the parameter $K_{sS}$ across all domain walls.  
The term $I_2$ in $G_{\theta_{sS},m,n}(\bm{r})$ is given by
    \begin{align}
        I_2 
        &= \frac{\Gamma_{{\rm em}}^{2}}{v_{{\rm dw}}\omega_{0}^{3}K_{sS}}\left[\Lambda-\delta_{n,0}\frac{\sin(\Lambda r)}{r}e^{-\omega_{0}\tau}\right] \; .
    \end{align}
Since $I_2$ arises from fully gapped modes,  this term approaches a constant, $\frac{\Lambda\Gamma^{2}}{v_{{\rm dw}}\omega_{0}^{3}K_{sS}}$, for $n=0$ in the long-distance and long-time limit, making it less significant than the $I_1$ term. 
Combining $I_1$ and $I_2$, we obtain  
    \begin{align}
        G_{\theta_{sS},m,n}(\bm{r})=\begin{cases}
\frac{v_{{\rm dw}}^{\prime}}{K_{sS}v_{{\rm dw}}}\ln\left|\frac{\tilde{r}}{\tilde{a}}\right|, & n=0\,,\\
\frac{v_{{\rm dw}}^{\prime}}{K_{sS}v_{{\rm dw}}}\left(-\gamma-\ln\frac{\tilde{a}}{L}\right)+\frac{\Lambda\Gamma_{{\rm em}}^{2}}{v_{{\rm dw}}\omega_{0}^{3}K_{sS}}\,, & n\neq0\,.
\end{cases}
    \end{align}

The correlation exponents, $K_{sS}^\prime$ and $(K_{sS}^\prime)^{-1}$, for $\phi_{sS}$ and $\theta_{sS}$ are thus modified by the electron-magnon coupling and exhibit the opposite behaviors, 
\begin{align}
    \frac{K_{sS}^\prime}{K_{sS}}
    = 
    \frac{1}{\sqrt{1-\Gamma_{{\rm em}}^{2}/v_{{\rm dw}}^{2}\omega_{0}^{2}}} \; ,
\end{align}
which is given in Eq.~(9) of the main text. 
Remarkably, when the electron-magnon coupling reaches the limit $\Gamma_{\rm em} \to v_{{\rm dw}}\omega_{0}$, the modified exponent shows a singularity, i.e. $K_{sS}^\prime \to \infty$.
This behavior influence  correlation functions and scaling dimensions of various operators.

\section{Spin relaxation rate}
\label{app:NMR}

In this section, we provide details for the derivation of the  relaxation rate $1/TT_1$ due to the exchange interaction, $H_{\rm{K}}$ of Eq.~(4) of the main text, of the conduction electrons and the local moments.
The  relaxation time $T_1$ measures the average spin-flip time by external perturbations ($H_{\rm{K}}$ in our work). By applying the Fermi's golden rule, the relaxation rate $1/T_1$  is related to the transverse spin susceptibility $\chi^R_{\perp}$ through~\cite{slichter-book}
\begin{align}
    \frac{1}{T_{1}T}=J_{K}^{2}a^{2}\frac{k_{B}}{\hbar^{3}}\lim_{\omega\to0}\frac{{\rm Im}\left[\chi_{\perp}^{R}(r=0,\omega)\right]}{\omega}\;,
    \label{eq:spin-relaxation-rate}
\end{align}
with  $\chi_{\perp}^{R}(r=0,t)=-i\left\langle T_{t}[s^{+}_m(r,t),\;s^{-}_m(r,t=0)]\right\rangle$,  $\chi_{\perp}^{R}(\omega)=\int dt e^{i\omega t}\chi_{\perp}^{R}(t)$, and $s^{\pm}_m=s^{x}_m\pm is^{y}_m$. 
Here, we keep only the interbranch scattering contributions, which dominate over the less relevant intrabranch ones, 
\begin{align}
    s^{\pm}_m (r) \to s^{\pm}_{\inter,m} (r)= \frac{1}{2}\left\lbrace O^{\pm}_{\inter,\rm{sdw}}(r) + \left[ O^{\mp}_{\inter,\rm{sdw}}\right]^\dagger(r) \right\rbrace\; ,
\end{align}
where we define the operator as
\begin{align}
    O_{\inter,{\rm sdw}}^{\pm}(r)  = & \frac{e^{-2iQ_{F}r}}{\pi a}\sum_{\delta}e^{i\left(\phi_{cS,m}\pm\delta\phi_{sA,m}\mp\theta_{sS,m}-\delta\theta_{cA,m}\right)} 
    \;.
\end{align}
Below, we investigate the $T>T_{{\rm hx}}$ and $T \ll T_{{\rm hx}}$ regimes separately.
 
\subsection{For $T>T_{{\rm hx}}$}

We now consider the temperature regime without the helical spin order, where all bosonic modes are gapless. Following the analysis in Ref.~\cite{Giamarchi2003} and Sec.~\ref{app:RKKY}, we calculate the transverse spin susceptibility
\begin{align}
    \chi_{\inter,\perp}^{R}(\omega)
    & = -\frac{2a}{v_{{\rm dw}}}\frac{\sin\left(\pi g_{\inter}^{x}\right)}{(\pi a)^{2}}\left(\frac{2\pi a k_B T}{\hbar v_{{\rm dw}}}\right)^{2g_{\inter}^{x}}   B\left(-\frac{i\hbar\omega}{2\pi k_B T}+g_{\inter}^{x};1-2g_{\inter}^{x}\right) \;.
\end{align}
To compute $1/T_{1}T$, we need to evaluate 
    \begin{align}
    \lim_{\omega\to0}\frac{{\rm Im}\left[\chi_{{\inter},\perp}^{R}(\omega)\right]}{\omega}=& \lim_{\omega\to0}\partial_{\omega}{\rm Im}\left[\chi_{{\inter},\perp}^{R}(\omega)\right] \nn
    =& \frac{\hbar^{2}}{2\pi^{2}k_{B}^{2}}\frac{\sin\left(\pi g_{{\inter}}^{x}\right)}{(\pi a)^{2}}\left(\frac{2\pi ak_{B}}{\hbar v_{{\rm dw}}}\right)^{2g_{{\inter}}^{x}}T^{2g_{{\inter}}^{x}-2}\Gamma\left(1-2g_{{\inter}}^{x}\right)\frac{\Gamma\left(1-2g_{{\inter}}^{x}\right)\Gamma\left(g_{{\inter}}^{x}\right)}{\Gamma\left(1-g_{{\inter}}^{x}\right)}\left[\Psi(g_{{\inter}}^{x})-\Psi(1-g_{{\inter}}^{x})\right] \; ,
\end{align}
with the digamma function, $\Psi(z)\equiv\left(1/\Gamma(z)\right)d\Gamma(z)/dz$. This leads to
    \begin{align}
        \frac{1}{T_{1}T}=\frac{J_{K}^{2}}{2\pi^{4}k_{B}\hbar}\sin(\pi g_{{\inter}}^{x})\left(\frac{2\pi ak_{B}}{\hbar v_{{\rm dw}}}\right)^{2g_{{\inter}}^{x}}T^{2g_{{\inter}}^{x}-2}\left\{ \Gamma\left(1-2g_{{\inter}}^{x}\right)\frac{\Gamma\left(1-2g_{{\inter}}^{x}\right)\Gamma\left(g_{{\inter}}^{x}\right)}{\Gamma\left(1-g_{{\inter}}^{x}\right)}\left[\Psi(g_{{\inter}}^{x})-\Psi(1-g_{{\inter}}^{x})\right]\right\} \;.
        \label{eq:spinrelax-above-Thx}
    \end{align}

\subsection{For $T\ll T_{{\rm hx}}$}
The spin helix generates an effective magnetic field, which gaps out half of   the electronic spectrum and therefore reduces the spin relaxation channels. Using the new boson representation, $\Phi^\eta_{\delta,m}$ and   $\Theta^\eta_{\delta,m}$ introduced in Section~\ref{app:Thx}, we obtain the following operator, 
\begin{align}
    O_{\inter,{\rm sdw}}^{\pm}(r)=\frac{e^{-2iQ_{F}r}}{\pi a}\sum_{\delta}e^{2i\Phi_{\delta}^{\mp}(r)}\;.
\end{align}
The corresponding transverse spin susceptibility, $\tilde{\chi}_{{\inter},\perp}$,  includes contributions from the $\Phi_{\delta}^{+}$ and $\Phi_{\delta}^{-}$ modes,
\begin{align}
    \tilde{\chi}_{{\inter},\perp}(\tau)&=\frac{-1}{4(\pi a)^{2}}\sum_{\delta}\left[\left\langle e^{2i\Phi_{\delta}^{-}(\tau)}e^{-2i\Phi_{\delta}^{-}(0)}\right\rangle   +\left\langle e^{-2i\Phi_{\delta}^{+}(\tau)}e^{2i\Phi_{\delta}^{+}(0)}\right\rangle\right]  \;.
    \end{align}

The contribution of the gapless $\Phi_{\delta}^{-}$ mode to $1/TT_1$ is given by 
    \begin{align}
        \frac{J_{K}^{2}}{4\pi^{4}k_{B}\hbar}\sin(\pi\bar{K}) \left[ 1-m_{2Q_F}(T) \right] \left(\frac{2\pi ak_{B}}{\hbar\tilde{v}_{{\rm dw}}}\right)^{2\bar{K}}T^{2\bar{K}-2}\left\{ \Gamma\left(1-2\bar{K}\right)\frac{\Gamma\left(1-2\bar{K}\right)\Gamma\left(\bar{K}\right)}{\Gamma\left(1-\bar{K}\right)}\left[\Psi(\bar{K})-\Psi(1-\bar{K})\right]\right\} \;, 
        \label{eq:spinrelaxT-Phim}
    \end{align}
showing a dependence $\propto \left[ 1-m_{2Q_F}(T) \right] T^{2\bar{K}-2}$.
Compared to Eq.~(\ref{eq:spinrelax-above-Thx}) for  $T>T_{\rm hx}$, we note several differences. 
First, due to the ordering of localized moments, $1/TT_1$ acquires a factor of $(1-m_{2Q_F})$ to account for the component of the localized spins that remain disordered and thus allow for spin-flip relaxation.  
Second, there is an overall prefactor of 1/2 to account for the remaining gapless electrons. 
Third, the  parameters depend on the modification from the helix, including $g_{{\inter}}^{x} \to \bar{K}$ and $v_{{\rm dw}} \to \tilde{v}_{{\rm dw}}$.

In addition, $1/TT_1$ receives contributions from the $\Phi_{\delta}^{+}$ modes, which are gapped and exhibit an exponentially suppressed relaxation rate,
    \begin{align}
        &\frac{J_{K}^{2}}{4\pi^{4}k_{B}\hbar}\left[ 1-m_{2Q_F}(T) \right]\left(\frac{2\pi ak_{B}}{\hbar \tilde{v}_{\rm dw}}\right)^{2\bar{K}} T^{2\bar{K}-2} {\rm Im}\Bigg\{ e^{i\pi \bar{K}}\frac{\Gamma(1-2\bar{K})\Gamma\left(-\frac{i\Delta (T)}{2\pi k_{B}T}+\bar{K}\right)}{\Gamma\left(-\frac{i\Delta(T)}{2\pi k_{B}T}+1-2\bar{K}\right)} \nn
        & \qquad \qquad \times\left[\Psi\left(-\frac{i\Delta(T)}{2\pi k_{B}T}+\bar{K}\right)-\Psi\left(-\frac{i\Delta(T)}{2\pi k_{B}T}+1-2\bar{K}\right)\right]\Bigg\} \;,
        \label{eq:spinrelaxT-Phip}
    \end{align}
with the temperature-dependent gap,
\begin{align}
    \Delta (T) = \tilde{\Delta}_a \left( \frac{J_K S m_{2Q_F}(T)}{\tilde{\Delta}_a} \right)^{1/(2-\bar{K})}\;.
\end{align}
We summarize the power-law behavior of the $1/T_1 T$, Eqs.~(\ref{eq:spinrelax-above-Thx}) and~(\ref{eq:spinrelaxT-Phim}) in the two temperature regimes, in Eq.~(10a) and Eq.~(10b) in the main text. Additionally, we present their temperature dependencies, along with the subdominant contribution in Eq.~\eqref{eq:spinrelaxT-Phip}, in Fig.~3 in the main text
and its inset.

\section{Paramagnetic spin susceptibility}
\label{app:susceptibility}

In this section, we examine the paramagnetic spin susceptibility arising from the electron subsystem, specifically contributions from small-momentum transfer ($q\sim 0$) at $T \ll T_{\rm hx}$. In the absence of interactions, this reduces to the Pauli susceptibility of a free electron gas. Incorporating interactions in the domain wall network here, this observable quantity not only reveals the spin ordering, but also interaction effects through renormalized system parameters.

To proceed, we examine the susceptibility from the response to an external magnetic field, characterized by the following Zeeman term,
\begin{align}
    H_{\rm h} = -\sum_{m}\int dr \bm{h}\cdot\bm{s}_{m}(r) = -h\sum_m\int dr \left[s_{m}^{z}(r)\cos\varphi + s_{m}^{x}(r)\sin\varphi\right]\;,
    \label{eq:H_zee}
\end{align}
where $\bm{h}= h \hat{\bm{n}}$ and $h=g\mu_{B}B$ with  $g$ denoting the Land\'e $g$-factor for domain wall modes, $\mu_B$  the Bohr magneton, and $B$ the magnetic field strength. The in-plane external field $B\hat{\bm{n}}$ forms an angle $\varphi$ with the $z$ axis (the domain wall direction),  and its unit vector is given by $\hat{\bm{n}}=(\sin\varphi,\; 0,\; \cos \varphi)$.

The area magnetization $M$ contribution  (magnetic moment per unit area)   from the electron spins is defined as 
\begin{align}
    M=g\mu_B \frac{ k_B T}{\hbar LL_{\perp}}\sum_{m}\int d\tau dr\left\langle \hat{\bm{n}}\cdot\bm{s}_{m}(r)\right\rangle = \frac{k_B T}{ LL_{\perp}}\frac{1}{Z}\frac{\partial Z}{\partial B}\;,
\end{align}
where $L_{\perp} \equiv N_{\rm dw}d$, and $Z$ denotes the partition function. Consequently, the (area) paramagnetic susceptibility, $\chi_u =\mu_0 \partial M/\partial B$ (with the permeability $\mu_0$), is given by 
\begin{align}
   \chi_{u} =  \frac{\mu_{0}\left(g\mu_{B}\right)^{2}}{\hbar d}\sum_{m}\sum_{\mu,\mu^{\prime}}\int d\tau dr  n^{\mu}n^{\mu^{\prime}}\mathcal{D}_{m}^{\mu\mu^{\prime}}(r,\tau) \;,
\end{align}
where $\mu\; , \mu^{\prime}\in \{x,y,z\}$ and $\mathcal{D}_{m}^{\mu\mu^{\prime}}(r,\tau)=\left\langle s_{m}^{\mu}(r,\tau)s_{m=0}^{\mu^{\prime}}(r=0,\tau=0)\right\rangle -\left\langle s_{m}^{\mu}(r,\tau)\right\rangle \left\langle s_{m=0}^{\mu^{\prime}}(r=0,\tau=0)\right\rangle $. Each component of $\mathcal{D}_{m}^{\mu\mu^{\prime}}(r,\tau)$ will be analyzed in detail below.

To proceed, we consider the effective action incorporating contributions from the electron subsystem, free magnons,  electron-magnon interaction, and the coupling to the rotating magnetic field from the spin helix. We start from the ones for the magnon-modified $\phi_{sS}$ and $\theta_{sS}$ fields upon integrating out the magnon fields, as shown in Section~\ref{app:Thx}.  
The resulting contribution $S_{\rm ee+em} \equiv  S_{\rm ee} + S_{\rm mag} + S_{\rm em}$ and the spin helix contribution $S_{{\rm hx}}$ are given by
\begin{align}
    \frac{S_{{\rm ee+em}}}{\hbar}& =   \frac{k_B T}{2\pi\hbar LN_{{\rm dw}}}\sum_{\bm{p},q_{\perp}}\left\{ 2iq\omega_{n}\theta_{cS,q_{\perp}}(\bm{p})\phi_{cS,-q_{\perp}}(-\bm{p})+\frac{q^{2}v_{{\rm dw}}}{K_{cS}(q_{\perp})}\left|\phi_{cS,q_{\perp}}(\bm{p})\right|^{2}+q^{2}v_{{\rm dw}}K_{cS}(q_{\perp})\left|\theta_{cS,q_{\perp}}(\bm{p})\right|^{2}\right. \nn
    & \quad +  2iq\omega_{n}\theta_{sS,q_{\perp}}(\bm{p})\phi_{sS,-q_{\perp}}(-\bm{p})+\frac{q^{2}v^\prime_{{\rm dw}}}{K^\prime_{sS}}\left|\phi_{sS,q_{\perp}}(\bm{p})\right|^{2}+q^{2}v^\prime_{{\rm dw}}K^\prime_{sS}\left|\theta_{sS,q_{\perp}}(\bm{p})\right|^{2} \nn
    & \quad +\sum_{\nu,P\neq cS,sS}\left. \left[2iq\omega_{n}\theta_{\nu P,q_{\perp}}(\bm{p})\phi_{\nu P,-q_{\perp}}(-\bm{p})+q^{2}v_{{\rm dw}} \left|\phi_{\nu P,q_{\perp}}(\bm{p})\right|^{2}+q^{2}v_{{\rm dw}}\left|\theta_{\nu P,q_{\perp}}(\bm{p})\right|^{2}\right]\right\}, 
    \nn[5pt]
    \frac{S_{{\rm hx}}}{\hbar}&\approx  \frac{k_B T }{2\pi \hbar L N_{{\rm dw}}}  \sum_{\bm{p},q_{\perp}} \Delta^{2} \left[\left|\phi_{cS,q_{\perp}} (\bm{p})\right|^{2}+\left|\theta_{cA,q_{\perp}} (\bm{p}) \right|^{2}+\left|\phi_{sA,q_{\perp}}(\bm{p})\right|^{2}+\left|\theta_{sS,q_{\perp}}(\bm{p}) \right|^{2}\right.\nn
    &  \quad  \left.  +2\phi_{cS,q_{\perp}}(\bm{p})  \theta_{sS,-q_{\perp}} (-\bm{p})  +2\theta_{cA,q_{\perp}}(\bm{p})  \phi_{sA,-q_{\perp}}(-\bm{p})  \right],
\end{align}  
where $v^\prime_{\rm dw}$ is defined in Eq.~(\ref{eq:vdw-prime}), $\Delta^{2}\equiv 2B_{{\rm hx}} /\hbar a $ 
is defined such that $\hbar \Delta \sqrt{v_{\rm dw}K_{cS}(q_\perp)}$  represents the magnitude of energy gap, and $K_{sS}^{\prime}$ given in Eq.~(9) of the main text. 

We evaluate the $\chi^{zx}_u$ component, which corresponds to the correlation between $s^z_m$ and $s^x_m$, as given by
\begin{align}
    \left\langle s_{m}^{z}(r,\tau)s_{m=0}^{x}(r =0,\tau=0)\right\rangle  
    & \sim\left\langle \left[\partial_{r}\phi_{sS,m}(r,\tau)\right]e^{i\left[\ell\phi_{sS,m=0}(r=0,\tau = 0 )-\theta_{sS,m=0}(r=0,\tau = 0 ) \right]}\right\rangle \nn
    & \qquad \times \left\langle e^{i\left[\ell\delta\phi_{sA,m=0}(r=0,\tau = 0 )-\delta\theta_{sA,m=0}(r=0,\tau = 0 )\right]}\right\rangle \;,
 \end{align}
where we  use  the $q\sim 0$ component of $s_{m}^{z}(r) = -\pi^{-1}\partial_{r}\phi_{sS,m}(r)$ in Eq.~(\ref{eq:s_z-phi_sS}) and $s_{m}^{x}(r)=\frac{1}{2\pi a}\sum_{\ell\delta}\cos\Big[\ell\phi_{sS,m}+\ell\delta\phi_{sA,m}-\theta_{sS,m}-\delta\theta_{sA,m}\Big]$. It is straightforward to obtain $\left\langle e^{i\left[\ell\delta\phi_{sA,m=0}(r=0,\tau = 0 )-\delta\theta_{sA,m=0}(r=0,\tau = 0 )\right]}\right\rangle = 0$, implying $\chi^{zx}_u = 0$. This result applies to all three temperature regimes. 

Next, we compute $\chi_u^{zz} (T \ll T_{\rm hx})$ and obtain
\begin{align}
    \chi^{zz}_u
     & = \frac{\mu_{0}\left(g\mu_{B}\right)^{2}\cos^2\varphi}{\hbar d}\int d\tau dr \left[\left\langle s_{m}^{z}(r,\tau)s_{m^{\prime}=0}^{z}( 0, 0)\right\rangle -\left\langle s_{m}^{z}(r,\tau)\right\rangle \left\langle s_{m^{\prime}=0}^{z}(0, 0)\right\rangle \right] \nn
     & = \frac{-\mu_{0}\left(g\mu_{B}\right)^{2}\cos^2\varphi}{\pi^{2}\hbar d}\left(\frac{k_B T}{\hbar LN_{{\rm dw}}}\right) \sum_{\bm{p},q_{\perp}}\sum_{\bm{p}^{\prime},q_{\perp}^{\prime}}\delta_{\bm{p},0}\delta_{q_{\perp},0}qq^{\prime}\Bigg[\left\langle \phi_{sS,q_{\perp}}(\bm{p})\phi_{sS,q_{\perp}^{\prime}}(\bm{p}^{\prime})\right\rangle -\left\langle \phi_{sS,q_{\perp}}(\bm{p})\right\rangle \left\langle \phi_{sS,q_{\perp}^{\prime}}(\bm{p}^{\prime})\right\rangle \Bigg]\;,
     \label{eq:chi_s}
\end{align}
Since  the subsequent analysis on $\chi_u^{zz}$ only involves the spin-symmetric sector (denoted by sS), the other sectors do not enter the following discussion. By integrating out  $\phi_{cS}$ and $\theta_{cS}$, the effective action, denoted as $S_{sS}$, is obtained as 
\begin{align}
    \frac{S_{sS}}{\hbar}&=\frac{k_B T}{2\pi\hbar LN_{{\rm dw}}}\sum_{\bm{p},q_{\perp}}\Bigg\{2iq\omega_{n}\theta_{sS,q_{\perp}}(\bm{p})\phi_{sS,-q_{\perp}}(-\bm{p})+\frac{q^{2}v^\prime_{{\rm dw}}}{K^\prime_{sS}}\left|\phi_{sS,q_{\perp}}(\bm{p})\right|^{2}\nn
   &  \qquad \qquad +\left[q^{2}v_{{\rm dw}}K_{sS}+\Delta^{2}-\frac{\Delta^{4}v_{{\rm dw}}K_{cS}(q_{\perp})}{\omega_{n}^{2}+v_{{\rm dw}}^{2}q^{2}+\Delta^{2}v_{{\rm dw}}K_{cS}(q_{\perp})}\right]\left|\theta_{sS,q_{\perp}}(\bm{p})\right|^{2}\Bigg\}\;.
   \label{eq:S_S}
\end{align}
Next, we integrate out the $\theta_{sS}$ field from Eq. (\ref{eq:S_S}) to derive the effective action that contains only $\phi_{sS}$. The corresponding propagator for $\phi_{sS}$ is then given by
\begin{align}
    \left\langle \phi_{sS,q_\perp}(\bm{p})\phi_{sS,-q_\perp}(-\bm{p})\right\rangle _{{\rm ee}+{\rm em}+{\rm hx}}=\frac{\pi\hbar LN_{{\rm dw}}/(k_B T)}{\frac{q^{2}v^\prime_{{\rm dw}}}{K^\prime_{sS}}+\frac{q^{2}\omega_{n}^{2}}{q^{2}v_{{\rm dw}}K_{sS}+\Delta^{2}-\frac{\Delta^{4}v_{{\rm dw}}K_{cS}(q_{\perp})}{\omega_{n}^{2}+v_{{\rm dw}}^{2}q^{2}+\Delta^{2}v_{{\rm dw}}K_{cS}(q_{\perp})}}}\;.
    \label{eq:phi_sS-prop_fin}
\end{align}
Applying Eqs. (\ref{eq:chi_s}) and (\ref{eq:phi_sS-prop_fin}), $\chi_u^{zz} (T\ll T_{\rm hx})$ reads
\begin{align}
    \chi_{u}^{zz} (T\ll T_{\rm hx}) = &  \frac{\mu_{0}\left(g\mu_{B}\right)^{2}\cos^2 \varphi}{\pi^{2}\hbar d}\left(\frac{k_B T}{\hbar LN_{{\rm dw}}}\right)\lim_{\omega_{n}\to0}\lim_{q,q_{\perp}\to0}q^{2}\left\langle \phi_{sS,q_{\perp}}(\bm{p})\phi_{sS,-q_{\perp}}(-\bm{p})\right\rangle \nn
    = &\frac{\mu_{0}\left(g\mu_{B}\right)^{2}\cos^2 \varphi }{\pi \hbar d}\lim_{\omega_{n}\to0}\left[ \frac{1}{\frac{v^\prime_{{\rm dw}}}{K^\prime_{sS}}+\frac{\omega_{n}^{2}}{\Delta^{2}-\frac{\Delta^{4}v_{{\rm dw}}\tilde{K}_{cS}(q_{\perp}=0)}{\omega_{n}^{2}+\Delta^{2}v_{{\rm dw}}\tilde{K}_{cS}(q_{\perp}=0)}}}\right] \nn
    =&\frac{\mu_{0}\left(g\mu_{B}\right)^{2}}{\pi \hbar v_{{\rm dw}} d} \frac{K_{sS}\cos^2 \varphi}{\left(1-\frac{\Gamma_{{\rm em}}^{2}}{\omega_{0}^{2}v_{{\rm dw}}^{2}}\right) +\tilde{K}_{cS}(q_\perp = 0)K_{sS}}\;,
    \label{eq:chi_zz-T-less-Thx}
\end{align}
where the appearance of $\tilde{K}_{cS}$ in Eq.~\eqref{eq:chi_zz-T-less-Thx}  above originates from the hybridization of $\theta_{sS}$ and $\phi_{cS}$  in $H_{\rm hx}$, induced by the helix ordering gap. The limits $\lim_{\omega_n \to 0}$ and $\lim_{q,q_\perp \to 0}$ do not commute and is thus not interchangeable here.

Subsequently, we compute $\chi_u^{xx}$ for $T\ll T_{\rm hx}$, which requires the computation of
\begin{align}
    \left\langle s_{m}^{x}(r,\tau)s_{m^{\prime}=0}^{x}( 0, 0)\right\rangle & =  \frac{1}{(2\pi a)^2}\sum_{\ell\ell^\prime\delta\delta^\prime} \left\langle \cos\left[\ell\phi_{sS,m}+\ell\delta\phi_{sA,m}-\theta_{sS,m}-\delta\theta_{sA,m}\right]  \right. \nn
    & \qquad \qquad \times \left. \cos\left[\ell^\prime\phi_{sS,m=0}+\ell^\prime \delta^\prime \phi_{sA,m=0}-\theta_{sS,m=0}-\delta^\prime \theta_{sA,m=0}\right]\right\rangle \nn
    & \propto  \left\langle e^{i\left[\theta_{sA,m}(r,\tau)-\theta_{sA,m=0}(r=0,\tau=0)\right]}\right\rangle\left\langle e^{i\left[\phi_{sS,m}(r,\tau)-\phi_{sS,m=0}(r=0,\tau=0)\right]}\right\rangle \;.
\end{align}
Since the relevant sine-Gordon term in $H_{\rm hx}$ stabilizes the fields $\phi_{sA}$ and $\theta_{sS}$, their conjugate fields $\theta_{sA}$ and $\phi_{sS}$ become disordered. Consequently, the correlation functions $\left\langle e^{i\left[\theta_{sA}(r,\tau) - \theta_{sA}(r^\prime, \tau^\prime)\right]}\right\rangle$ and $\left\langle e^{i\left[\phi_{sS}(r,\tau) - \phi_{sS}(r^\prime, \tau^\prime)\right]}\right\rangle$ decay exponentially. Therefore, $\chi_u^{xx}$ is exponentially small for temperatures $T \ll T_{\rm hx}$. We thus retain the dominant contribution and reach
\begin{align}
    \chi_{u} (T\ll T_{\rm hx}) \approx \chi_{u}^{zz} (T\ll T_{\rm hx}) =\frac{\mu_{0}\left(g\mu_{B}\right)^{2}}{\pi \hbar v_{{\rm dw}} d} \frac{K_{sS}\cos^2 \varphi}{\left(1-\frac{\Gamma_{{\rm em}}^{2}}{\omega_{0}^{2}v_{{\rm dw}}^{2}}\right) +\tilde{K}_{cS}(q_\perp = 0)K_{sS}}\;.
    \label{eq:chi_u-full-T-less-Thx}
\end{align}
In the $T\to 0$ limit, where magnons are absent,  $\chi_u(T\to 0 )$ can be straightforwardly obtained by taking $\Gamma_{\rm em}\to0$ from Eq.~\eqref{eq:chi_u-full-T-less-Thx}, leading to
\begin{align}
    \chi_{u} (T\to 0) =\frac{\mu_{0}\left(g\mu_{B}\right)^{2}}{\pi \hbar v_{{\rm dw}} d} \frac{K_{sS} \cos^2 \varphi }{1 +\tilde{K}_{cS}(q_\perp = 0)K_{sS}} \;.
    \label{eq:chi_u-full-T-0}
\end{align}

Above the helix ordering temperature $T > T_{\rm hx}$, the spin rotational symmetry is preserved. Consequently, the full susceptibility $\chi_u$ can be obtained by simply aligning the magnetic field along the $z$ direction, as expressed by
\begin{align}
    \chi_{u} (T>T_{\rm{hx}}) = \frac{\mu_{0}\left(g\mu_{B}\right)^{2}}{\pi^{2}\hbar d}\lim_{q,q_{\perp}\to0}\lim_{\omega_{n}\to0}\left(q^{2}\frac{\pi v_{{\rm dw}}K_{sS}}{v_{{\rm dw}}^{2}q^{2}+\omega_{n}^{2}}\right) = \mu_{0}\left(g\mu_{B}\right)^{2} \frac{K_{sS}  }{\pi\hbar v_{{\rm dw}}d}\;.
    \label{eq:chi_u-T-greater}
\end{align}
This result  can be directly obtained  by setting the $\tilde{K}_{cS}$ term to zero in the denominator of Eq.~\eqref{eq:chi_u-full-T-0}. This is because the hybridization between $\theta_{sS}$ and $\phi_{cS}$ vanishes for $T > T_{\rm hx}$, as the helix-ordering gap is destroyed. Alternatively, Eq.~\eqref{eq:chi_u-T-greater} can be reproduced by shifting the boson field, $\tilde{\phi}_{sS,q_{\perp}=0}(r)=\phi_{sS,q_{\perp}=0}(r)+\frac{K_{sS}N_{{\rm dw}}h}{\hbar v_{{\rm dw}}}r$~\cite{Giamarchi2003}. 

In Table~I of the main text, we summarize the paramagnetic spin susceptibility for the three temperature regimes, in Eqs.~\eqref{eq:chi_u-full-T-0}--\eqref{eq:chi_u-T-greater}.
Equation~\eqref{eq:chi_u-full-T-less-Thx} reveals that the paramagnetic susceptibility in the completely ordered phase, $ \chi_{u} (T\to 0) $, is reduced compared with $\chi_{u} (T > T_{\rm hx})$. This reduction  captures the partially gapped electron spectra and is controlled by the SLL parameters through  $\tilde{K}_{cS}(q_\perp =0)$, exhibiting experimental tunability.   In the noninteracting limit $K_{cS}, \; K_{sS}\to1$, the  susceptibility becomes exactly half of its value in the $T>T_{\rm hx}$ regime. 

Next, we give remarks on    
the order of taking the limits $q\to 0$ and $\omega_n\to 0$.
These two limits in general do not commute. In calculating  $\chi_u(T>T_{\rm hx})$, taking the $\omega_n \to 0$ limit first reproduces the result using the alternative method discussed earlier~\cite{Giamarchi2003}. In contrast, in  calculating  $\chi_u(T\ll T_{\rm hx})$, taking the limits  in the reversed order leads to  physically reasonable results for both $ T \ll T_{\rm hx} $ and $ T\to 0$.
Namely, the susceptibility in the   ordered phase  shows a reduction,  reflecting the partial gap in the electron spectrum.   
 
To assess these the predicted observables, here we estimate their overall scale, $\frac{\mu_{0}\left(g\mu_{B}\right)^{2}}{\pi \hbar v_{{\rm dw}} d}$; see Table~I of the main text.   For  $\theta = 0.5\degree$ and   $U_{\rm ee}/\hbar v_{\rm dw} = 23$, we have $d\approx 2.45 \times 10^{-8}$ m and $v_{\rm dw} \approx 10^{-5}$ m/s~\cite{HCWang-TBG}. While   experimental   $g$-factor of correlated domain wall modes  is still lacking, we note that $g \approx 2.12$ was reported in a half-filled twisted double bilayer graphene~\cite{TDBG-Cheng-2020}. This motivates our choice   $g = 2$ and lead to the molar paramagnetic susceptibility $\chi_{u,\rm{mol}} = \chi_u\cdot m_{\rm mol}/\rho_{\rm mass}$ on the order of 
\begin{align} \label{Eq:chi_estimation}
    \chi_{u,\rm{mol}} \sim \frac{m_{\rm mol}}{\rho_{\rm mass}} \times O\left( \frac{\mu_{0}\left(g\mu_{B}\right)^{2}}{\pi \hbar v_{{\rm dw}} d} \right)  \approx 4.2 \times 10^{-12}\;\rm{m}^3/\rm{mol}\;,
\end{align}
with the molar mass $m_{\rm mol}=12 \; \rm{g/mol}$  and (area) mass density $\rho_{\rm mass} = 1.53\times 10^{-3}\;\rm{g/m}^2$ for twisted bilayer graphene.  In the more widely used CGS unit, the molar susceptibility is expressed as $\chi_{u,\rm{mol}}=3.34\times  10^{-7} \;\rm{emu}/\rm{mol}$. Here, we also present the mass susceptibility, defined as $\chi_{u,\rm{mass}}= \chi_u/\rho_{\rm mass} = 2.78\times 10^{-8}\;\rm{emu}/\rm{g}$.

\let\temp\addcontentsline
\renewcommand{\addcontentsline}[3]{} 


\putbib[DW-TBG-bib-Supp]
\let\addcontentsline\temp

\end{bibunit}

\end{document}